\documentclass[10pt,conference]{IEEEtran}
\IEEEoverridecommandlockouts

\usepackage[utf8]{inputenc}
\usepackage[T1]{fontenc}
\usepackage{etex}
\usepackage{graphicx}
\usepackage{fleqn}
\usepackage{cite}
\usepackage{physics}
\usepackage{siunitx}
\usepackage{amsmath,amssymb,amsfonts}
\usepackage{algorithmic}
\usepackage{graphics}
\usepackage{textcomp}
\usepackage{tikz}
\usepackage{relsize}
\usepackage{color}
\usetikzlibrary{arrows, arrows.meta}
\usepackage{comment}
\usepackage{xcolor}
\usepackage{filecontents}
\usepackage{float}
\usetikzlibrary{arrows.meta,patterns}
\usetikzlibrary{shapes.misc}
\usepackage{graphics}
\usepackage{pict2e,color}
\usepackage{bbm}
\usepackage{subcaption}
\usepackage{nccmath}
\usepackage{mathtools}
\usetikzlibrary{fit}
\usepackage{nccmath}
\usepackage{pgfplots}
\usepackage{lipsum}
\usepackage{stackengine}
\usepackage[capitalize]{cleveref}
\usepackage[ruled,vlined]{algorithm2e}
\usepackage{amsthm}
\usepackage{balance}
\usepackage{color,soul}
\usepackage{dsfont}
\usepackage{arydshln}

\usepackage{booktabs}

\usepackage{acro}

\usepackage[colorinlistoftodos,prependcaption,textsize=footnotesize]{todonotes}
\usepackage{regexpatch}

\usepackage[utf8]{inputenc}
\usepackage[T1]{fontenc}
\usepackage{etex}
\usepackage{graphicx}
\usepackage{fleqn}
\usepackage{cite}
\usepackage{physics}
\usepackage{siunitx}
\usepackage{amsmath,amssymb,amsfonts}
\usepackage{algorithmic}
\usepackage{graphics}
\usepackage{textcomp}
\usepackage{tikz}
\usepackage{relsize}
\usepackage{color}
\usetikzlibrary{arrows, arrows.meta}
\usepackage{comment}
\usepackage{xcolor}
\usepackage{filecontents}
\usepackage{float}
\usetikzlibrary{arrows.meta,patterns}
\usetikzlibrary{shapes.misc}
\usepackage{graphics}
\usepackage{pict2e,color}
\usepackage{bbm}
\usepackage{subcaption}
\usepackage{nccmath}
\usetikzlibrary{fit}
\usepackage{nccmath}
\usepackage{pgfplots}
\usepackage{lipsum}
\usepackage{stackengine}
\usepackage[capitalize]{cleveref}
\usepackage[ruled,vlined]{algorithm2e}
\usepackage{amsthm}
\usepackage{balance}
\usepackage{color,soul}
\usepackage{dsfont}
\usepackage{acro}
\usepackage{booktabs}
\usepackage[colorinlistoftodos,prependcaption,textsize=footnotesize]{todonotes}
\usepackage{regexpatch}
\usepackage{enumitem}
\usepackage{tasks}
\setul{0.5ex}{0.3ex}
\setulcolor{blue}
\def\BibTeX{{\rm B\kern-.05em{\sc i\kern-.025em b}\kern-.08em
    T\kern-.1667em\lower.7ex\hbox{E}\kern-.125emX}}

\makeatletter
\xpatchcmd{\@todo}{\setkeys{todonotes}{#1}}{\setkeys{todonotes}{inline,#1}}{}{}
\makeatother
\usepackage{cleveref}

\crefname{equation}{}{}
\Crefname{equation}{}{}
\crefname{section}{Sect.}{Sects.}
\Crefname{section}{Sect.}{Sects.}

 \newtheorem{Lemma}{Lemma}

\def\1{\mathbf{1}}
\def\0{\mathbf{0}}

\def\Ebb{\mathbb{E}}
\def\Rbb{\mathbb{R}}

\def\Rbb{\mathbb{R}}

\def\Binom{\text{Binom}}

\def\e{\text{e}}

\def\argmin{\text{argmin}}

\begin{document}

\author{\IEEEauthorblockN{Karim S. Elsayed  and Amr Rizk }
\IEEEauthorblockA{\textit{Leibniz University Hannover, L3S, Germany} 
}
}

\title{Balancing Quantum Memories in Asymmetric Repeaters for High-Fidelity Entanglement Distribution}
\maketitle

\begin{abstract}
At the core of the quantum Internet lie quantum repeaters that enable remote end-to-end entanglement generation.
Fundamentally, the entanglement generation rate and fidelity of quantum repeaters constitute the bottleneck for end-to-end performance.
To achieve high rates, quantum repeaters employ quantum memory multiplexing. 
In a high-rate standard repeater, each memory \textit{sequentially} generates an entanglement with its neighboring nodes and then applies entanglement swapping.
This, however, results in low fidelity due to decoherence of the first-formed entanglement in the sequential generation process. 

By allocating different numbers of memories to \textit{simultaneously} form entanglements with the left and right adjacent nodes, quantum repeaters reduce high waiting times and achieve high fidelity. 
In such a repeater, a \textit{mismatch} problem arises due to the difference between the probabilistic number of generated entanglements on both sides.
Consequently, some entanglements remain stored until opposite entanglements are available.
The \textit{mismatch problem} reduces the repeater rate and particularly the entanglement fidelity.
In this paper, we consider the mismatch problem in an \textit{asymmetric repeater} with different distances to its adjacent nodes.
To mitigate the mismatch problem, we derive a dynamic optimal memory allocation.
Under the optimal allocation, we derive statistical lower bounds on the achievable rate and fidelity.
We demonstrate that the optimal allocation significantly improves the fidelity while maintaining a comparable rate to the standard repeater.
In contrast, our results show that fixed memory allocation may be detrimental to the fidelity.

\end{abstract}

\vspace{-5pt}
\section{Introduction and Background}

\label{sect: intro}

Quantum repeaters lie at the core of the quantum Internet, enabling communication between two remote quantum nodes~\cite{inesta2023optimal,azuma2023quantum}.
Due to the exponential decay of qubits with distance through quantum channels~\cite{low_loss_optical_fiber}, repeaters form short-distance link-level connections and  extend the link-level entanglements with its adjacent nodes through entanglement swapping~\cite{inesta2023optimal}.
Typically, distant remote end nodes require several repeaters~\cite{inesta2023optimal}.
Several protocols exist in the literature in terms of entanglement swapping order and decisions~\cite{gu2024fendi,Ghaderibaneh}.
A common goal of such protocols is to reduce the end-to-end entanglement generation duration to achieve high fidelity and rates. 
To further enhance the rates, repeaters use multiplexing with multiple quantum memories that store generated entanglements~\cite{collins2007multiplexed,abruzzo2014finite,kunzelmann2025multiplexed}.
Essentially, end-to-end entanglement rates and fidelity are limited by rates and fidelities of each repeater in the path. 
Particularly, the fidelity decays exponentially with the number of repeaters~\cite{dur1999entanglement_purification,gu2024fendi}. 

Assuming  that the repeater resides between two nodes, in the standard repeater architecture model~\cite{rozpkedek2019near}, every quantum memory is responsible for \textit{sequentially} generating entanglements with the left and right nodes of the repeater. 
Once both entanglements are available, the repeater applies entanglement swapping. 
While the standard repeater rate is reasonably \textit{high} and not hardware demanding, its fidelity is relatively \textit{low}.
This stems from the sequential entanglement generation, where the fidelity of the first-generated entanglement decays in memory until the successful generation of an opposite entanglement.

To improve fidelity, several works consider a repeater architecture, which simultaneously generates entanglements with both adjacent nodes~\cite{kunzelmann2025multiplexed,lee2022quantum,Twosley_ideal_Qu_switch,vasantam2022throughput}. 
Such a repeater allocates different quantum memories to form entanglements with the left and right nodes.
The quantum memories \textit{simultaneously} attempt to form entanglements with their assigned node.
After each entanglement generation attempt, the repeater matches the successful entanglements from both sides and performs entanglement swapping~\cite{lee2022quantum}.
Because of simultaneous generation, the waiting time of entanglements is significantly shorter than that of the standard repeater, resulting in higher fidelity.
This comes at the expense of \textit{a lower rate, due to the mismatch} between the number of successfully generated entanglements with the left and right nodes~\cite{lee2022quantum}.
Observe that the number of successful entanglements with the left and right nodes is random, specifically, binomially distributed.
Such randomness stems from the inherent decay and possible loss of qubits in quantum channels~\cite{low_loss_optical_fiber,elsayed2024trade}.
Hence, some entanglements remain unmatched and wait for the next generation attempts.
Importantly, while the mismatch results in a loss in the repeater rate, its effect on the fidelity is significant.
This stems from the exponential decay of the fidelity with the waiting time of the unmatched entanglements~\cite{dur_standard}.

In this paper, we consider an asymmetric quantum repeater with different distances from the left and right nodes.
Our goal is to tackle the mismatch problem to optimize the repeater fidelity and rate. 
To this end, we analytically derive the optimal memory allocation in terms of the repeater-node distances, the number of unmatched entanglements and the total number of quantum memories.

Recent work focuses on the analysis of multiplexed repeaters considering simultaneous entanglement generation~\cite{vasantam2022throughput,Twosley_ideal_Qu_switch,Inside_quantum_repeater}.
The works in~\cite{Twosley_ideal_Qu_switch,vasantam2022throughput} consider the analysis of a quantum switch, i.e., multi-port repeater.
Specifically, the work in~\cite{Twosley_ideal_Qu_switch} analyzes quantum switches assuming an idealized fluid model. 
Under a fluid model, the authors in~\cite{vasantam2022throughput} derive the arrival rates resulting in switch stability.
In~\cite{Inside_quantum_repeater}, the authors derive the fidelity of a multiplexed repeater under purification. 
Our work differs from these works as we optimize the memory allocation for which we analytically provide statistical bounds.
Similarly, the work in~\cite{lee2022quantum} proposes an optical physical realization of the repeater.
The authors consider a symmetric repeater with an equal number of memories allocated to both sides and analytically derive a lower bound on the repeater rate.
In contrast, we consider a dynamic memory allocation to mitigate the entanglement generation mismatch for \textit{asymmetric} repeaters.

We summarize our contributions as: (i) We analytically derive the optimal mismatch-dependent memory allocation, which minimizes the expected number of unmatched entanglements. (ii) Under the optimal allocation, we derive analytical bounds on the expected fidelity and expected rate. (iii) As the optimal allocation condition may not be satisfied for specific repeater parameters, we devise a hard-cutoff allocation regime. (iv) Our numerical evaluations verify the superiority of the optimal allocation in terms of expected fidelity with negligible rate loss compared to the standard repeater.

\label{sect: related work}
\vspace{-5pt}

\section{Model and Problem statement}
\vspace{-5pt}
\label{sec: model}

\begin{figure}[t]
    \centering
    \begin{subfigure}{0.99\linewidth}
    \centering
    \includegraphics[width=0.8\linewidth]{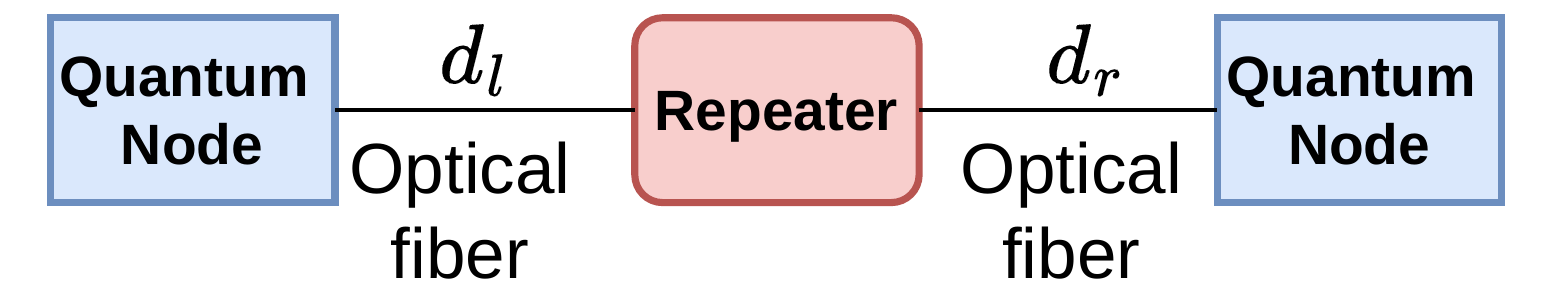}
    \vspace{-5pt}
    \caption{}
    \label{fig:Model_a}
    \end{subfigure}
    \begin{subfigure}{0.99\linewidth}
    \centering
    \includegraphics[width=1\linewidth, trim={1em 1em 2em 0em}, clip]{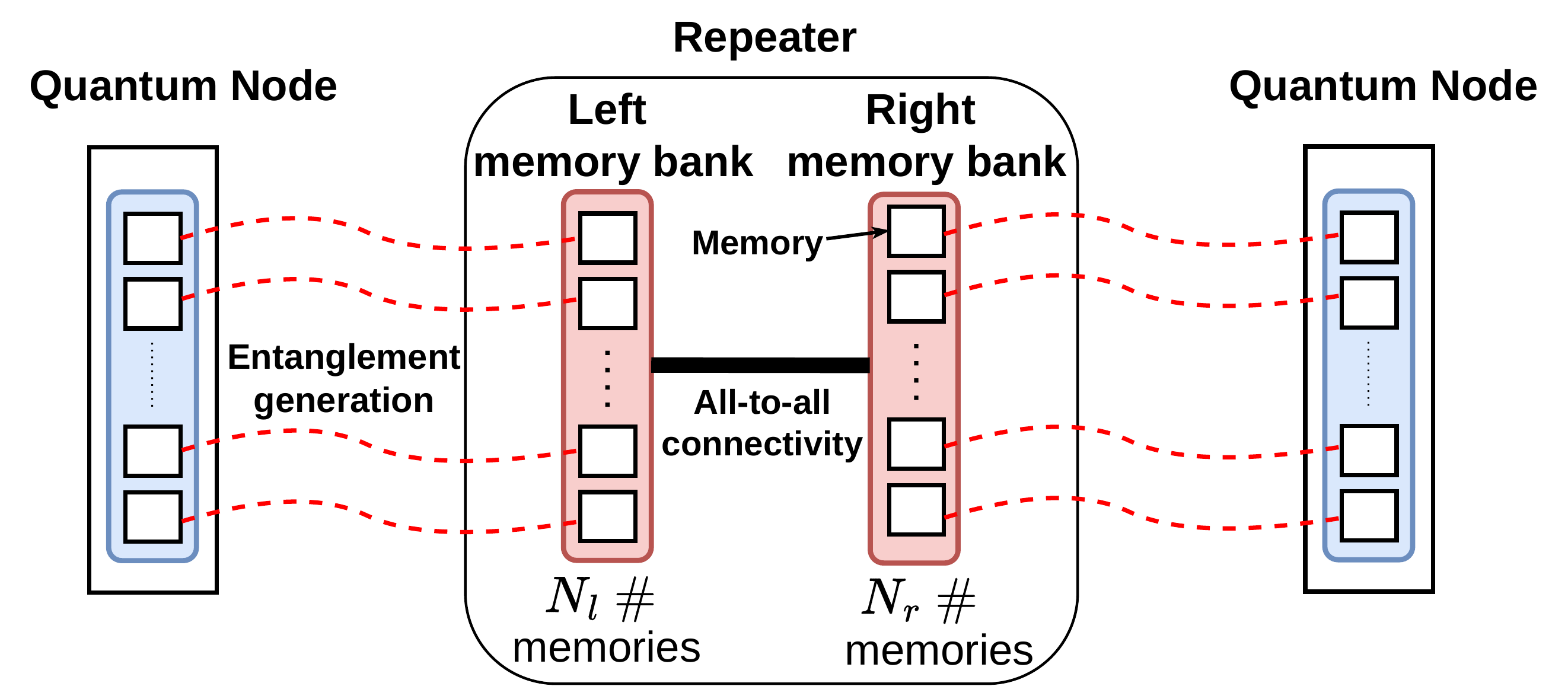}
    \caption{}
    \label{fig:Model_b}
    \end{subfigure}
    \caption{(a) \textbf{System-level view:} Asymmetric quantum repeater connecting two quantum nodes with different distances $d_l\neq d_r$. 
    Quantum nodes may be end nodes or quantum repeaters.
    (b) \textbf{Schematic illustration of the considered repeater architecture:}
    The repeater contains $N$ memories dynamically allocated between the left and right sides to form entanglements with neighboring nodes.
    The number of memories $N_l$ allocated to the left is depicted by the left memory bank, while $N_r$ memories are allocated to the right.
    The memories in the left and right banks \textit{simultaneously} attempt to generate entanglements with left and right adjacent nodes, respectively, which is depicted by the dashed red curves.
     The thick line connecting the banks represents all-to-all connectivity between any two memories in both banks.
    The depiction does not necessarily reflect the physical implementation of the repeater.
    }
    \label{fig:Model}
    \vspace{-15pt}
\end{figure}

We consider an optical quantum repeater with multiplexed quantum memories as illustrated in~\cref{fig:Model}. The repeater connects two quantum nodes, which may be end nodes or other repeaters.  
Specifically, the repeater contains $N$ quantum memories, each of which can be freely connected to either of the two nodes on the left and right side through a quantum channel, i.e., optical fiber.
The repeater sides are, in general, asymmetric, i.e., the distances between the repeater and its left node $d_l$ and its right node $d_r$ may not be equal.  
The repeater shares end-to-end entanglements between its connected nodes in a round of \textit{two phases}: (i) entanglement generation and (ii) matching and swapping.
Before each round, the repeater assigns $N_l$ and $N_r$ memories to establish entanglements with the left and right node, respectively.
We depict the memory assignment by the left and right memory banks in~\cref{fig:Model_b}.

In the entanglement generation phase, memories in each bank \textit{simultaneously} attempt to generate entanglements with their assigned node.
As a result, $X_l$ and $X_r$ entanglements are successful with the left and right nodes, respectively, and get stored in the corresponding memory banks.
In the second phase, the repeater \textit{matches} each pair of stored entanglements from both memory banks, then applies entanglement \textit{swapping} to form end-to-end entanglements\footnote{We use end-to-end entanglements to refer to the entanglement formed between the repeater adjacent nodes} between the left and right nodes.

\paragraph{Entanglement Generation Model} A variety of photon-mediated entanglement generation protocols may be considered, e.g., Barrett-Kok~\cite{barrett2005efficient}, Duan-Kimble~\cite{duan2004scalable} or Moehring–Madsen (only compatible with trapped ions)~\cite{moehring2004experimental}.
In such protocols, two spin-photon entanglements resulting from electron excitation in two connected repeater-node memories are swapped, forming a spin-spin entanglement between the electrons.
Several types of memories may be considered, particularly Nitrogen Vacancy (NV) centers~\cite{schirhagl2014nitrogen} or trapped ions~\cite{moehring2004experimental}, due to their relatively long coherence times.
Considering NV centers, spin-spin entanglements between electrons are further swapped to the nuclear spins with longer coherence times. 
Importantly, we consider channel multiplexing, where entanglements on each side of the repeater are generated over a shared physical channel.
Specifically, at each repeater side, the photons mediating entanglements from different memories are transmitted together over the same physical channel using, e.g., spectral or temporal channel multiplexing.
Hence, only one physical channel, i.e., an optical fiber, is required to connect the nodes at each side of the repeater.
The entanglement generation duration between the repeater and each of its adjacent nodes is \textit{constant}, dominated by the photon trip time, i.e., the time of the photon transmission. 
The photon trip time to the left or the right nodes is given by $\tau_l = d_l/c$ and $\tau_r = d_r/c$, respectively, where $c$ is the speed of light.
As photons may be absorbed by the optical fiber, entanglement generation between two memories is probabilistic. 
Accordingly, entanglement generation for each side is a Bernoulli trial with success probability $p_l=\e^{-\sigma d_l}$ and $p_r=\e^{-\sigma d_r}$, respectively.
Here, the parameter $\sigma$ is the decay rate. 
Hence, the number of successfully generated entanglements $X_l$ stored in the left bank with $N_l$ memories and the number of successfully generated entanglements $X_r$ stored in the right bank with $N_r$ memories are binomially distributed.

\paragraph{Matching and Swapping Model}
To form an end-to-end entanglement, the repeater matches left bank memories containing $X_l$ entanglements and right bank memories containing $X_r$ entanglements to perform entanglement swapping.
Specifically, matching involves physically connecting every two memories from the left and right banks.
One physical realization proposed in~\cite{lee2022quantum} is a  Mach-Zehnder interferometer (MZI) based switch, which provides all-to-all connectivity between \textit{NV centers}, thereby enabling a physical connection between any two memories in the repeater.
Additionally, the switch enables connection from any memory in the repeater to any memory in either the left or the right connected nodes, hence supporting free assignment of memories to the left or the right bank. 
Alternatively, trapped-ion repeaters typically contain several ions confined in a single trap, which naturally supports entanglement swapping between any two ions~\cite{dhara2022multiplexed}. 
Recall that each ion acts as one quantum memory.
Notably, the matching and swapping phase occurs locally within the repeater with significantly shorter local distances between the memories than repeater-node distances $d_r$ and $d_l$.
As a result, the repeater round duration is dominated by the entanglement generation phase duration.

After each repeater round, some entanglements either in the left or the right bank remain unmatched, due to the mismatch between the number of successfully generated entanglements $X_l$ and $X_r$ stored in the left and right bank, respectively.
Unmatched entanglements typically remain stored in the next repeater rounds until entanglements from the opposite bank become available.
Let $\alpha(t)$ denote the number of unmatched entanglements stored at the beginning of round $t$.
At the beginning of round $t+1$, the number of unmatched entanglements is given by~\cite{lee2022quantum}
\vspace{-5pt}
\begin{equation}
\label{eq:mismatch_t}
    \alpha(t+1)=\alpha(t)+X_l(t)-X_r(t)\,.
\vspace{-5pt}
\end{equation}
Here, the sign of $\alpha$ identifies the location of unmatched entanglements, i.e, $\alpha >0$ when entanglements remain in the left bank, while $\alpha <0$ when entanglements remain in the right bank.
As previously explained, the number of generated entanglements $X_l(t)$ and $X_r(t)$ are binomially distributed as
\vspace{-5pt}
\begin{align}
\label{eq:ent_gen_binom}
    X_l(t) & \sim
    \begin{cases}
        \Binom(N_l(t)-\alpha(t),p_l), \quad \alpha(t) \geq 0 \,, \\
        \Binom(N_l(t),p_l), \quad \alpha(t) < 0 \,,
    \end{cases}
     \nonumber \\
    X_r(t) &\sim
    \begin{cases}
        \Binom(N_r(t),p_r), \quad \alpha(t) \geq 0 \,, \\
        \Binom(N_r(t)+\alpha(t),p_r), \quad \alpha(t) < 0 \,,
    \end{cases}
\vspace{-5pt}
\end{align}
where $\Binom(n,p)$ is the binomial distribution with $n$ trials and probability of success $p$. 
Here, the left and right memory bank sizes $N_l(t)$ and $N_r(t)$, respectively, include the memories containing unmatched entanglements $\alpha(t)$.
Accordingly, $N_l(t)-\alpha(t)$ memories are free in the left bank when $\alpha(t) \geq 0$, with which the repeater attempts to generate entanglement with the left node. 
Observe that $N_l(t)-\alpha(t)$ is the number of trials of the Binomial distribution of $X_l$ in the first line in~\cref{eq:ent_gen_binom}.
Similarly, $N_r(t)+\alpha(t)$ memories are free in the right bank when $\alpha(t)<0$, which is the number of trials of the Binomial distribution of $X_r$ in~\cref{eq:ent_gen_binom} when $\alpha(t)<0$.

\begin{figure}[t]
    \centering
    \includegraphics[width=0.8\linewidth, trim={5.5em 1em 5.5em 2em}, clip]{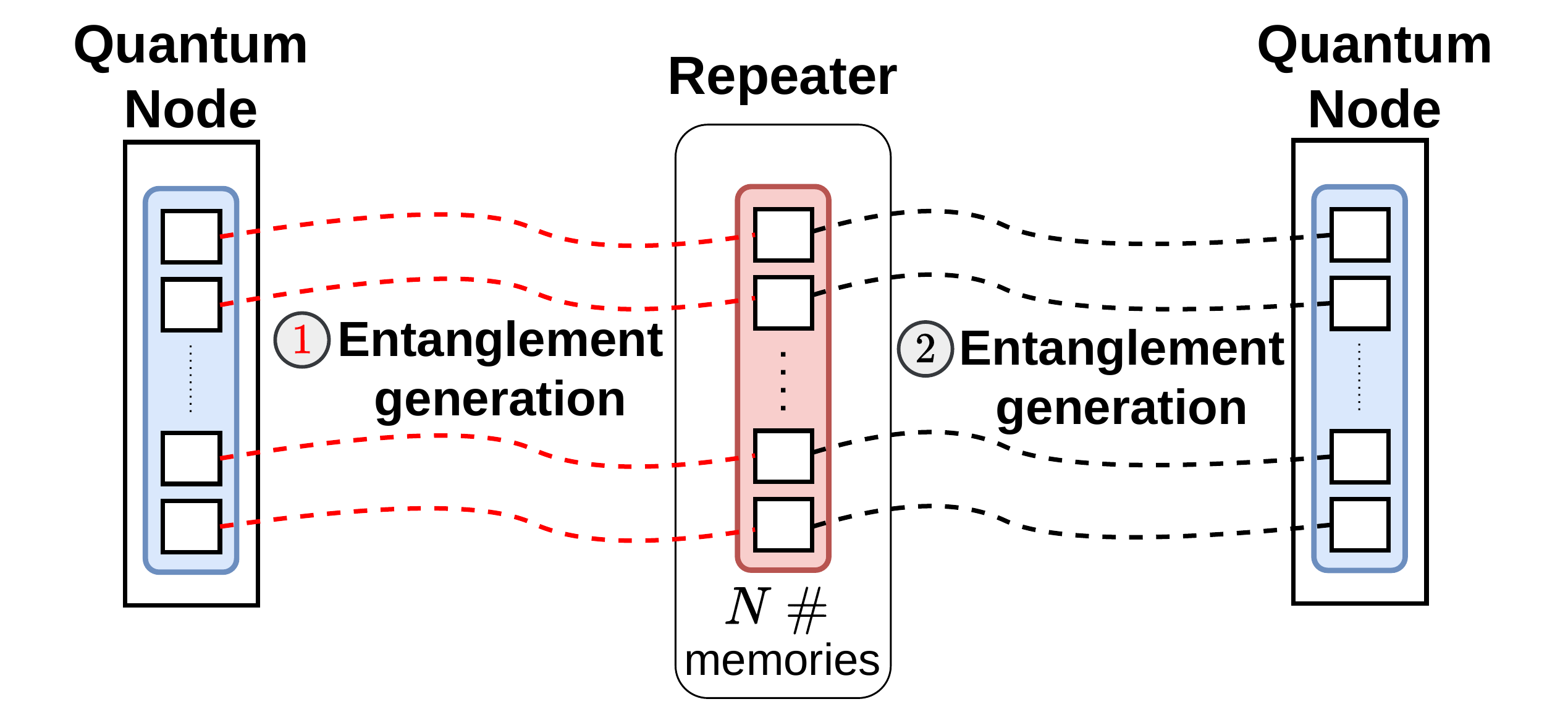}
    \caption{Schematic illustration of the standard repeater architecture. The repeater contains $N$ memories, each of which sequentially forms an entanglement with the two adjacent nodes. The red and black dashed curves depict the sequential entanglement generation with the left and right nodes, respectively.}
    \label{fig:std_rep_model}
    \vspace{-15pt}
\end{figure}

The key advantage of simultaneous entanglement generation on both sides of the repeater is the significantly\textit{ higher fidelity} achieved compared to a standard repeater.
In a standard repeater, every quantum memory first generates an entanglement with one side~\cite{rozpkedek2019near}, as depicted in~\cref{fig:std_rep_model}. 
Once successful, the same memory generates an entanglement with the opposite side, then applies entanglement swapping. 
NV centers are suitable for such repeaters since they can store two simultaneous spin-spin entanglements in the nuclear and the electron spins.
Specifically, the electron in the NV center is responsible for forming the subsequent entanglements with the two opposite nodes. 
The nucleus spins store the first entanglement until the electron forms an entanglement with the opposite node. 
Consequently, the fidelity of the first entanglements decays significantly, particularly under low coherence times and long distances. 
Note that we similarly consider channel multiplexing for the standard repeater, where only a single physical channel is required at each side of the repeater.

While simultaneous generation in the considered repeater improves fidelity, it is limited by the \textit{mismatch} between the entanglements successfully generated in the two memory banks.
Such limitations stem from the waiting time of unmatched entanglements $\alpha$ in the repeater, which may significantly increase, particularly in asymmetric repeaters or dynamically changing links under fixed memory allocation. 
We consider the depolarization noise model assuming entanglements are affected by isotropic noise. 
Hence, the state of entanglements is a Werner state, which decays with the age $A$ as~\cite{dur_standard}
\vspace{-5pt}
\begin{equation}
\label{eq:fid_dec}
    F(A)=0.25+\left(F_0-0.25\right) \e^{-A\tau_{\text{round}}/t_c}\,, 
\vspace{-5pt}
    \end{equation}
where $t_c$ is the coherence time. 
Note that $A$ is the discrete age counter that increases in time steps of repeater rounds $\tau_{\text{round}}$.
The fidelity resulting from the entanglement swapping of two Werner states with fidelities $F_1$ and $F_2$ is~\cite{dur1999entanglement_purification}
\vspace{-5pt}
\begin{equation}
\label{eq:swap}
    F_{\text{swap}}=F_1F_2+ \frac{1}{3} (1-F_1)(1-F_2)\,.
\vspace{-5pt}
\end{equation}
Note that we do not consider possible fidelity losses due to hardware operations, e.g., Bell state measurements, as they are constant under any memory allocation regime.
Additionally, compared to the standard repeater, the considered repeater architecture requires more gates, specifically to implement all-to-all connectivity between each memory pair, for example, using a Mach-Zehnder interferometer-based switch as previously discussed.
However, as demonstrated in~\cite{lee2022quantum}, such fidelity loss is outweighed by the fidelity loss due to depolarization noise when the link distances are not significantly short.
Notably, corresponding decay coefficients can be directly incorporated in~\cref{eq:swap} as given in~\cite[Equations (3-4)]{gu2024fendi}.

Our goal is to minimize the mismatch $\alpha$ through the optimal allocation of the quantum memories, specifically 
\vspace{-5pt}
\begin{equation}
\label{eq:mismatch_min}
        \min_{N_l(t),N_r(t)} \Ebb[\abs{\alpha(t+1)}]\,.
\vspace{-5pt}
        \end{equation}
Here, $N_l(t),N_r(t)$ are the optimal bank sizes at round $t$ that minimize the unmatched entanglements remaining in the next round $\alpha(t+1)$.
This minimization has a two-fold advantage: It optimizes fidelity and improves the entanglement rate.
Essentially, the rate improvement stems from the resulting increase in the expected number of matched entanglements per unit time.

\vspace{-5pt}
\section{Optimal Balancing and Analytical Evaluation}
\label{sect:approach}
In this section, we analytically derive the optimal memory allocation of asymmetric repeaters, which minimizes the repeater mismatch.
Additionally, we derive lower bounds on the entanglement rate and fidelity.

Next, we derive the expected stationary rate R of the repeater in terms of the expected stationary mismatch $\Ebb\left[\abs{\alpha}\right]$.
We calculate the expected rate as the ratio between the expected number of entanglements per repeater round and the round duration, i.e., $R={\Ebb[M]}/{\tau_{\text{round}}}$. 
Recall from~\cref{sec: model} that the entanglement generation duration dominates the repeater round duration.
As the distances from the left and right nodes are not equal, the round duration is the largest of the entanglement generation durations, expressed as
\vspace{-5pt}
\begin{equation}
\label{eq:round_time}
    \tau_{\text{round}}=\max(\tau_l,\tau_r), 
\vspace{-5pt}
\end{equation}
where the durations $\tau_l=d_l/c$ and $\tau_r=d_r/c$ are the entanglement generation durations with the left and right nodes, respectively, and $c$ is the speed of light.

Next, the number of matched entanglements in round $t$ starting with $\alpha(t)$ stored entanglements due to the mismatch in the previous round is
\vspace{-5pt}
\begin{align}
    M(t)=\frac{1}{2} \left( X_l(t)+X_r(t)+\abs{\alpha(t)}-\abs{\alpha(t+1)}\right)\,,
    \vspace{-5pt}
\end{align}
where $X_l(t)$ and $X_r(t)$ are the successfully generated entanglement in the left and right banks, respectively, which are defined in~\cref{eq:ent_gen_binom}.
Here, $X_l(t)+X_r(t)+\abs{\alpha(t)}$ is the total number of available entanglements in both memory banks after the entanglement generation phase, and $\abs{\alpha(t+1)}$ remains unmatched.
Under stationarity, we derive the expected number of matched entanglements as 
\vspace{-5pt}
\begin{equation}
\label{eq:E_matched}
    \Ebb[M]=\frac{1}{2}\left(\Ebb[X_l]+\Ebb[X_r]\right)\,.
\vspace{-5pt}
\end{equation}
Note that $X_l$ and $X_r$ depend on $\alpha$ from~\cref{eq:ent_gen_binom}.
We exactly derive $\Ebb[M]$ for asymmetric repeaters in the Appendix~\ref{app:asymmetric_rate} as
\vspace{-5pt}
\begin{align}
\label{eq:n_ent_general}
    \Ebb[M]&=\frac{1}{2}\big(p_l\Ebb[N_l] +p_r\Ebb[N_r]  \nonumber \\
    &-\left(p_l \Ebb[\abs{\alpha},\alpha \geq 0]+p_r \Ebb[\abs{\alpha}, \alpha < 0]\right)\big)\,,
\vspace{-5pt}
    \end{align}
which we bound from below as
\begin{equation}
\label{eq:n_ent_bound_general}
    \Ebb[M]\geq \frac{1}{2}\left(p_l \Ebb[N_l] +p_r \Ebb[N_r] - \max{(p_l,p_r)} \Ebb[\abs{\alpha}]\right)\,.
\end{equation}
Recall that $p_l$ and $p_r$ are the probabilities of entanglement generation success with the left and right nodes, respectively. Equations~\cref{eq:n_ent_bound_general,eq:n_ent_general} demonstrate the improvement in the rate when minimizing the mismatch $\Ebb[\abs{\alpha}]$.
Notably, the rate is maximized for symmetric repeaters, i.e., $p_l=p_r$, where the expected number of matched entanglements follows from~\cref{eq:n_ent_general} as $\Ebb[M]= \frac{1}{2} p_l\left(N- \Ebb[\abs{\alpha}]\right).$
Note that this equation coincides with the analysis of symmetric repeaters in~\cite{lee2022quantum}.
\vspace{-5pt}
\subsection{Optimal Memory Allocation}
Now, we derive the optimal memory allocation $N_l(t)$ and $N_r(t)$ to the left and right memory banks, respectively, which minimizes the repeater mismatch according to~\cref{eq:mismatch_min}.
It is known that the median $\text{med}({\alpha})$ minimizes the absolute error given as $\text{med}({\alpha})= \underset{a\in \Rbb}{\argmin}\ \Ebb[\abs{\alpha-a}]$~\cite{degroot2005optimal}.
Accordingly, the optimal memory allocation must fulfill the condition $\text{med}({\alpha})=0$ to minimize the expectation $\Ebb[\abs{\alpha}]$.
Under this condition, obtaining analytically explicit expressions for the optimal bank sizes $N_l(t)$ and $N_r(t)$ is challenging. 
This stems from the complexity of deriving the stationary distribution of $\abs{\alpha}$ for an arbitrary total number of memories $N$, even under fixed memory allocation.

Alternatively, we derive the optimal conditional allocation on the number of unmatched entanglements $\alpha(t)$ in round $t$ by driving the mean of $\alpha(t+1)$ in the next round $t+1$ as 
\vspace{-5pt}
\begin{equation}
\label{eq:zero_mean_cond}
        \Ebb[\alpha(t+1)|\alpha(t)]=0\,.
\end{equation}
This condition is approximately optimal in the limit of the number of memories $N$, where the relative error between the mean and the median of $\alpha(t+1)$ diminishes.
For illustration, note that under the Central Limit Theorem (CLT), the binomial distribution is well approximated by a Gaussian distribution, with diminishing approximation error in the limit~\cite{billingsley2017probability}. 
The conditional expectation and conditional median of the unmatched entanglements from~\cref{eq:mismatch_t} are
\vspace{-5pt}
\begin{align}
       \Ebb[ \alpha(t+1)|\alpha(t)]=\alpha(t)+\Ebb[X_l(t)-X_r(t)|\alpha(t)]\,, \nonumber \\
       \text{med}[ \alpha(t+1)|\alpha(t)]=\alpha(t)+\text{med}[X_l(t)-X_r(t)|\alpha(t)]\,.
       \vspace{-5pt}
\end{align}
Recall from~\cref{eq:ent_gen_binom} that for a given $\alpha(t)$, the number of entanglement generation successes $X_l(t)$ and $X_r(t)$ are binomially distributed.
Under the Central Limit Theorem, the distributions of $X_l(t)$ and $X_r(t)$ are approximated by Gaussian distributions. 
Consequently, the distribution of the difference $X_l(t)-X_r(t)$ is also Gaussian. 
As a result, the mean and median of the unmatched entanglements are equal, i.e., $\Ebb[ \alpha(t+1)|\alpha(t)]=\text{med}[ \alpha(t+1)|\alpha(t)]$.

In the lemma below, we derive the \textit{unique} allocation that enforces the zero-mean condition in~\cref{eq:zero_mean_cond}. 
Together with the asymptotically achieved zero-median condition, this yields an asymptotically optimal allocation for minimizing the expected mismatch.
\vspace{-5pt}
\begin{Lemma}
\label{lemm:optimal_alloc}
    Given the total number of quantum memories $N$ and the probabilities of entanglement generation $p_l$ and $p_r$ with the left and right nodes, respectively. The right and left bank sizes in round $t$ under the zero-mean condition $\Ebb[\alpha(t+1)|\alpha(t)]=0$ are 
\vspace{-5pt}
        \begin{align*}
        N_r(t)&=\left\lfloor \frac{p_l}{p_l+p_r}\left(N+\alpha(t) \frac{1-g(\alpha(t))}{p_l}\right)\right\rfloor\,, \nonumber \\
        N_l(t)&=N-N_r(t)\,,
        \vspace{-5pt}
    \end{align*}
    respectively with $g(\alpha(t))=1_{\{\alpha(t) \geq 0\}}p_l+1_{\{\alpha(t) < 0\}}p_r$.
\end{Lemma}
The proof of Lem.~\ref{lemm:optimal_alloc} is in the Appendix~\ref{app:optimal_alloc}. 
This allocation \textit{balances the expected entanglement generation rates with the left and the right node}.
Observe that as $\alpha(t)$ increases, the right bank size increases, effectively compensating for the imbalance on the left side.
We note that we limit the left bank size as $N_r(t) \in [1,N-1]$.

\vspace{-5pt}
\subsection{Lower Bound on Repeater Rate}
Next, we give a lower bound on the entanglement rate. 
\vspace{-5pt}
\begin{Lemma}
 \label{Lemm:rate_bound}
     Given the total number of quantum memories $N$ and the probabilities of entanglement generation $p_l$ and $p_r$ with the left and right nodes, respectively. The expected number of matched entanglements $\Ebb[M]$ under the memory allocation from Lemma~\ref{lemm:optimal_alloc} is bounded from below as
  \vspace{-5pt}
  \begin{align}
     \label{eq:M_boundd}
         \Ebb[M] & \geq \frac{1}{2} \left(2 \eta N- \max(p_l,p_r) \sqrt{\eta (2-p_r-p_l) N} \right)\,, 
  \vspace{-5pt}
     \end{align}
     where $\eta={p_r p_l}/({p_l+p_r})$.
 \vspace{-5pt}
 \end{Lemma}
The proof of Lem.~\ref{Lemm:rate_bound} is in the Appendix~\ref{app:rate_bound}. 
As the repeater round time $    \tau_{\text{round}}=\max(\tau_l,\tau_r)$ is constant, the rate $R$ is bounded from below as  $R\geq { \Ebb[M]}/{\tau_{\text{round}}}$.
The rate bound in~\cref{eq:M_boundd} approaches the standard repeater rate in the limit  as the second term changes with $N$ as $\mathcal{O}(\sqrt{N})$, while the first term is $\mathcal{O}({N})$.
Notably, the expected rate of the standard asymmetric repeater is bounded as
\vspace{-5pt}
\begin{equation}
\label{eq:std_bound}
    R_{\text{std}}=N/\Ebb[\tau_\text{round,std}]\geq{\eta N}/{\tau_{\text{round}}}\,,
\vspace{-5pt}
\end{equation}
where $\tau_\text{round,std}$ is the standard repeater round time.
Recall that every memory in the standard repeater forms an end-to-end entanglement by sequentially generating an entanglement with the two connected nodes. 
Hence, the expected round time $\tau_\text{round,std}$ is the \textit{sum of the entanglement generation times with the left and right nodes}. 
Precisely, the entanglement generation duration is then geometrically distributed.
As a result $\Ebb[\tau_\text{round,std}]=\tau_l/p_l+\tau_r/p_r$ and the rate for $N$ memories is $N/\Ebb[\tau_\text{round,std}]$, which results in the bound in~\cref{eq:std_bound}.
Recall that the durations $\tau_l$ and $\tau_r$ are the entanglement generation attempt duration with the left and right nodes, respectively.
Note that the duration $\tau_{\text{round}}$ is the round duration in the considered repeater, given in~\cref{eq:round_time} is different from $\tau_\text{round,std}$.

\vspace{-5pt}
\subsection{Lower Bound on the Fidelity}
\label{sec:fid_LB}
Next, we derive a lower bound on the fidelity of the end-to-end entanglements. 
The fidelity decay below the initial fidelity $F_0$ originates from the waiting time of the unmatched entanglements $\alpha$ in any of the banks. 
Hence, we first derive a bound on the age of the unmatched entanglements. 
Consider a number of unmatched entanglements $\alpha(t)>0$ residing in the left memory banks at time $t$. 
We assume the router prioritizes older unmatched entanglements in the matching phase, i.e., First-in First-out matching.
The age of each entanglement in $\abs{\alpha(t)}$ is bounded by the age of the last matched one, which we denote as $\delta$.
Accordingly, we bound the expected age as
  \vspace{-5pt}
\begin{equation}
    \Ebb[A|\alpha(t)=q,q>0]\leq \Ebb[\delta(q)]\,.
  \vspace{-5pt}
\end{equation}
Precisely, the age $\delta(q)$ is the time until the right bank generates $q$ entanglements.
Hence, $\delta(q)$ is a hitting time, defined as
  \vspace{-5pt}
\begin{equation}
    \delta(q)=\text{inf}\big\{u : S_u=\sum_{i=1}^u X_r(i)\geq q \big\}. 
  \vspace{-5pt}
\end{equation}
Here, the random variable $X_r(i)$ is the number of entanglements generated at time step $i$ in the right bank.
According to the Additive drift theorem~\cite{he2004study}, the expected age $\Ebb[\delta(q)]$ is bounded as $\Ebb[\delta(q)] \leq q/\beta$ if the expected drift $\Ebb[X_r(i)]\geq \beta$ for all $i$.  
We know from Lem.~\ref{lemm:optimal_alloc} that the size of the right bank when $\alpha(t)>0$ is at least ${N_r(t)={p_lN}/({p_l+p_r}) }$.
Accordingly, the expected binomial drift $\Ebb[X_r(t)] \geq p_r N_r(t)$ for all $t$, and the bound $\beta={p_l p_r N}/({p_l+p_r}) =\eta N$.
Consequently, the expected age is bounded as
\begin{equation*}
    \Ebb[A|\alpha(t)=q,q>0]\leq {q}/{\eta N}\,.
\end{equation*}
Similarly, $\Ebb[A|\alpha(t)=q,q<0]\leq {-q}/{\eta N}\,.$
Combining both expressions, we obtain
\vspace{-5pt}
\begin{equation}
\label{eq:F_LB}
    \Ebb[A|\alpha(t)=q]\leq{\abs{q}}/{\eta N}\,.
\vspace{-5pt}
\end{equation}
Taking the expectation with respect to $\alpha(t)$ and under stationarity, the age is bounded as
\begin{equation}
     \Ebb[A] \leq \frac{\Ebb[\abs{\alpha}]}{\eta N} \leq \frac{\sqrt{\eta (2-p_r-p_l) N}}{\eta N}:=\mathcal{B}_A\,.
     \label{eq:F_LB_another}
     \vspace{-5pt}
\end{equation}
Here, we use the Cauchy-Schwarz inequality: $\Ebb[\abs{\alpha}]\leq \sqrt{\Ebb[\alpha^2]}$.
Note that as $ \Ebb[\alpha]=0$, $\Ebb[\alpha^2]=\text{Var}[\alpha]$, where we obtain the upper bound: $\text{Var}[\alpha]\leq \eta (2-p_r-p_l) N$ 
by substituting \eqref{eq:E_N} into \eqref{eq:var_alph} from the Appendix~\ref{app:rate_bound}.

Next, we derive the fidelity bound of the unmatched entanglements, which decay according to~\cref{eq:fid_dec} as
\vspace{-5pt}
\begin{align}
\label{eq:F_LB_1}
    \Ebb[F]&=0.25+\left(F_0-0.25\right) \Ebb\left[\e^{-A\tau_{\text{round}}/t_c}\right]\,, \nonumber \\
    &\geq 0.25+\left(F_0-0.25\right) \e^{-\mathcal{B}_A\tau_{\text{round}}/t_c}:=\mathcal{B}_F\,,
\vspace{-5pt}
\end{align}
where $F_0$ is the initial fidelity and $t_c$ is the coherence time.
The second line uses Jensen's inequality.
As any unmatched entanglement with non-zero age gets matched with a fresh entanglement once available, we bound the fidelity after swapping using~\cref{eq:swap} as 
\vspace{-5pt}
\begin{equation}
      \Ebb[F_{\text{swap}}]\geq F_0\mathcal{B}_F+ \frac{1}{3} (1-F_0)(1-\mathcal{B}_F)\,. 
      \vspace{-5pt}
  \end{equation}
As $F_{\text{swap}}$ is increasing in $F$, the lower bound of the unmatched fidelity $\Ebb[F]$ results in a lower bound after swapping $\Ebb[F_{\text{swap}}]$.
Note that we assume the possible fidelity loss in one round due to the asymmetric entanglement generation durations is negligible.

\vspace{-5pt}
\subsection{Optimal Allocation Violation and Hard-Cutoff Regimes}
\label{sec:Alloc_viol}
The optimal allocation zero-mean condition on the expected unmatched entanglements $\Ebb[\alpha(t+1)|\alpha(t)]=0$, may not be fulfilled if there is not a sufficient total number of memories that satisfy the allocation in Lem.~\ref{lemm:optimal_alloc}.
Observe that the optimal allocation $N_r(t)$ is not bounded in $[1,N]$, especially for large $\alpha$ and small probabilities.
For illustration, consider the success probabilities $p_l=2p_r$.
For zero unmatched entanglement $\alpha(t)=0$, under the optimal allocation in Lem.~\ref{lemm:optimal_alloc}, the repeater should assign twice as many memories to the right bank as the left one.
As a result, at least $N=3$ number of quantum memories are required to minimize the mismatch.  
Hence, for $N=2$, the zero-mean optimal condition is not fulfilled.
We refer to such a case as optimal allocation violation.
Notably, this violation is dependent on the number of unmatched entanglements, which is inherent to the allocation from Lem.~\ref{lemm:optimal_alloc}.
This means that the violation may occur only for certain values of unmatched entanglements exceeding a violation threshold, which we identify as
\vspace{-5pt}
\begin{align}
\label{eq:viol}
    N_r(t)&< \abs{\alpha(t)}\,, \quad \text{when  } \alpha(t) <0\,,\nonumber \\ 
    N_l(t)&< \abs{\alpha(t)}\,, \quad \text{when  } \alpha(t)>0. 
\vspace{-5pt}
\end{align}
As the memory bank sizes $N_r(t)$ and $N_l(t)$ include the memories with unmatched entanglements, any allocation requiring fewer memories than the available stored unmatched entanglements implies optimal allocation violation.
To always enforce the zero-mean optimal condition $\Ebb[\alpha(t+1)|\alpha(t)]=0$ for a fixed total number of memories $N$, we propose a hard-cutoff regime.
In that regime, we drop the extra unmatched entanglements above the violation threshold.
Inserting the memory bank sizes from Lem.~\ref{lemm:optimal_alloc} into~\cref{eq:viol}, we obtain the unmatched number of entanglement thresholds as
\vspace{-5pt}
\begin{align}
\label{eq:viol_1}
   \abs{\alpha(t)}> p_l N /(p_l+1)= \alpha^-_{\text{thr}}  \,, \quad \text{when  } \alpha(t) <0\,,\nonumber \\
     \abs{\alpha(t)}>  p_r N /(p_r+1)=\alpha^+_{\text{thr}}\,, \quad \text{when  } \alpha(t) >0\,,
\vspace{-5pt}
 \end{align}
By dropping the excess unmatched entanglements, we optimize the fidelity while sacrificing the lowest possible rate. 

\captionsetup[subfigure]{%
  justification=centering,
  singlelinecheck=true,
skip=1pt
}
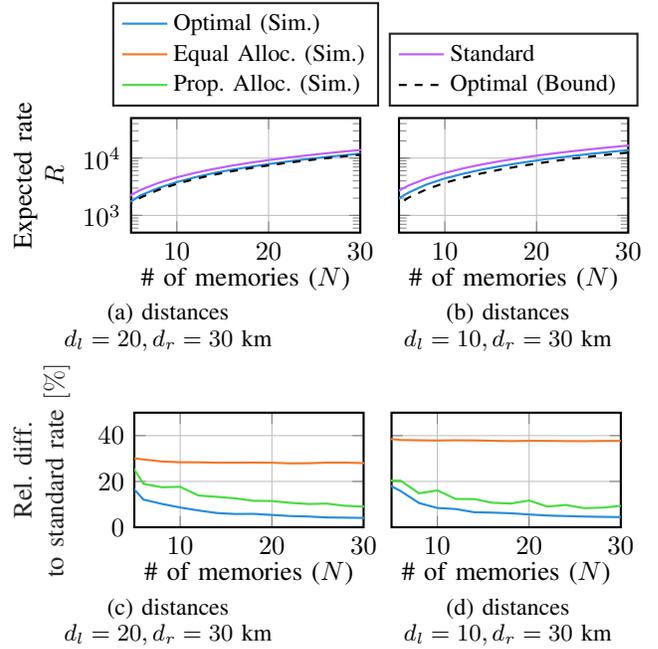
\begin{figure}[t]
\centering
\begin{subfigure}[b]{0.49\linewidth}
  \centering
%
%
\definecolor{mycolor1}{rgb}{0.14900,0.54900,0.86600}%
\definecolor{mycolor2}{rgb}{0.96000,0.46600,0.16000}%
\definecolor{mycolor3}{rgb}{1.00000,0.90900,0.39200}%
\definecolor{mycolor4}{rgb}{0.75200,0.36000,0.98400}%
\definecolor{mycolor5}{rgb}{0.28600,0.85800,0.25000}%
\definecolor{mycolor6}{rgb}{0.85098,0.85098,0.85098}%
\definecolor{mycolor7}{rgb}{0.07059,0.07059,0.07059}%
\begin{tikzpicture}

\begin{axis}[%
line width=0.8pt, 
ymode=log,
width=1.2in,
height=0.6in,
scale only axis,
xmin=5,
xmax=30,
xlabel style={font=\color{black}},
xlabel={\# of memories ($N$)},
xlabel style={yshift=0.2cm},
ymin=500,
ymax=50000,
ylabel style={font=\color{black},align=center},
ylabel={Expected rate \\ $R$},
axis background/.style={fill=white},
legend style={legend cell align=left, align=left, draw=white!15!black},
xmajorgrids,
ymajorgrids,
legend style={legend cell align=left, align=left},
legend style={at={(0.5,1.1)}, anchor=south, legend columns=1},
 legend style={font=\small}
]
\addplot [color=mycolor1, mark options={solid, mycolor1}]
  table[row sep=crcr]{%
2	533.56\\
4	1372.48\\
6	2221.1\\
8	3011.78\\
10	3819.24\\
12	4638.86\\
14	5468.32\\
16	6275.02\\
18	7055.28\\
20	7874.82\\
22	8700.94\\
24	9511.04\\
26	10344.14\\
28	11148.66\\
30	11960.08\\
32	12790.66\\
34	13589.28\\
36	14405.94\\
38	15208.16\\
40	16032.22\\
40	16039.32\\
60	24221.58\\
80	32391.82\\
100	40605.5\\
120	48838.2\\
140	57016.36\\
160	65311.6\\
180	73523.46\\
200	81709.6\\
};
\addlegendentry{Optimal (Sim.)}
\addplot [color=mycolor2, mark options={solid, mycolor2}]
  table[row sep=crcr]{%
2	529.36\\
};
\addlegendentry{Equal Alloc. (Sim.)}

\addplot [color=mycolor5,  mark options={solid, mycolor5}]
  table[row sep=crcr]{%
2	530.06\\
};
\addlegendentry{Prop. Alloc. (Sim.)}

\addplot [color=mycolor4, mark options={solid, mycolor4}]
  table[row sep=crcr]{%
2	923.680959673810\\
4	1847.36191934762\\
6	2771.04287902142\\
8	3694.72383869523\\
10	4618.40479836904\\
12	5542.08575804285\\
14	6465.76671771666\\
16	7389.44767739046\\
18	8313.12863706427\\
20	9236.80959673808\\
22	10160.4905564119\\
24	11084.1715160857\\
26	12007.8524757595\\
28	12931.5334354333\\
30	13855.2143951071\\
40	18473.6191934762\\
50	23092.0239918452\\
60	27710.4287902142\\
70	32328.8335885833\\
80	36947.2383869523\\
90	41565.6431853214\\
100	46184.0479836904\\
110	50802.4527820594\\
120	55420.8575804285\\
130	60039.2623787975\\
140	64657.6671771666\\
150	69276.0719755356\\
160	73894.4767739047\\
170	78512.8815722737\\
180	83131.2863706427\\
190	87749.6911690118\\
200	92368.0959673808\\
};

\addplot [color=black, dashed,  mark options={black, mycolor4}]
  table[row sep=crcr]{%
2	566.67478754167\\
4	1287.45083929925\\
6	2033.58034986626\\
8	2792.83377643147\\
10	3560.47410854807\\
12	4334.07191886506\\
14	5112.18401610019\\
16	5893.87008129476\\
18	6678.47696666941\\
20	7465.52911930157\\
22	8254.66753514764\\
24	9045.61330377692\\
26	9838.14462236547\\
28	10632.0816540363\\
30	11427.2761874177\\
40	15418.3692238368\\
50	19428.2159511897\\
60	23451.3840047731\\
70	27484.6463107754\\
80	31525.9002520845\\
90	35573.6853458662\\
100	39626.9382234589\\
110	43684.8561274169\\
120	47746.8153950574\\
130	51812.3200601886\\
140	55880.9679953287\\
150	59952.4277897753\\
160	64026.4224746363\\
170	68102.717768611\\
180	72181.1133983487\\
190	76261.4365640934\\
200	80343.5369360827\\
};
\end{axis}

\end{tikzpicture}%
  \caption{distances $d_l=20, d_r=30 \text{ km}$}
  \label{fig:rates_1}
\end{subfigure}
\hfill
\begin{subfigure}[b]{0.49\linewidth}
\hspace{10pt}
 {
%
%
\definecolor{mycolor1}{rgb}{0.14900,0.54900,0.86600}%
\definecolor{mycolor2}{rgb}{0.96000,0.46600,0.16000}%
\definecolor{mycolor3}{rgb}{1.00000,0.90900,0.39200}%
\definecolor{mycolor4}{rgb}{0.75200,0.36000,0.98400}%
\definecolor{mycolor5}{rgb}{0.28600,0.85800,0.25000}%
\definecolor{mycolor6}{rgb}{0.85098,0.85098,0.85098}%
\definecolor{mycolor7}{rgb}{0.07059,0.07059,0.07059}%
\begin{tikzpicture}

\begin{axis}[%
line width=0.8pt, 
ymode=log,
width=1.2in,
height=0.6in,
scale only axis,
xmin=5,
xmax=30,
xlabel style={font=\color{black}},
xlabel={\# of memories ($N$)},
ymin=500,
ymax=50000,
ylabel style={font=\color{black}},
xlabel style={yshift=0.2cm},
yticklabels={},
axis background/.style={fill=white},
legend style={legend cell align=left, align=left, draw=white!15!black},
xmajorgrids,
ymajorgrids,
legend style={legend cell align=left, align=left},
legend style={at={(0.5,1.1)}, anchor=south, legend columns=1},
 legend style={font=\small},
]
\addplot [color=mycolor1, forget plot,mark options={solid, mycolor1}]
  table[row sep=crcr]{%
2	573.32\\
4	1598.26\\
6	2484.5\\
8	3469.32\\
10	4422.34\\
12	5329.86\\
14	6298.98\\
16	7210.74\\
18	8134\\
20	9082.42\\
22	10035.2\\
24	10971.44\\
26	11909.18\\
28	12842.9\\
30	13774.68\\
32	14716.28\\
34	15670.3\\
36	16604.26\\
38	17551.1\\
40	18501.62\\
40	18499.64\\
60	27922.04\\
80	37373.74\\
100	46871.36\\
120	56334.92\\
140	65848.82\\
160	75317.38\\
180	84834.04\\
200	94426.88\\
};
\addplot [color=mycolor4, mark options={solid, mycolor4}]
  table[row sep=crcr]{%
2	1103.84226694070\\
4	2207.68453388130\\
6	3311.52680082200\\
8	4415.36906776270\\
10	5519.21133470340\\
12	6623.05360164400\\
14	7726.89586858470\\
16	8830.73813552540\\
18	9934.58040246610\\
20	11038.4226694067\\
22	12142.2649363474\\
24	13246.1072032881\\
26	14349.9494702288\\
28	15453.7917371694\\
30	16557.6340041101\\
40	22076.8453388135\\
50	27596.0566735168\\
60	33115.2680082202\\
70	38634.4793429236\\
80	44153.6906776269\\
90	49672.9020123303\\
100	55192.1133470337\\
110	60711.3246817371\\
120	66230.5360164404\\
130	71749.7473511438\\
140	77268.9586858472\\
150	82788.1700205505\\
160	88307.3813552539\\
170	93826.5926899573\\
180	99345.8040246607\\
190	104865.015359364\\
200	110384.226694067\\
};
\addlegendentry{Standard }

\addplot [color=black, dashed, mark options={solid, black}]
  table[row sep=crcr]{%
2	467.791778449548\\
4	1223.29843749321\\
6	2026.14129076681\\
8	2853.48716121917\\
10	3696.49180995313\\
12	4550.61937281624\\
14	5413.17533844573\\
16	6282.40408362658\\
18	7157.08614830888\\
20	8036.33365105293\\
22	8919.47631309067\\
24	9805.99339449386\\
26	10695.4707781342\\
28	11587.5726994361\\
30	12482.022440247\\
40	16982.5016415066\\
50	21517.9949354485\\
60	26078.3598214911\\
70	30657.571189061\\
80	35251.7033453067\\
90	39858.0294946606\\
100	44474.5642738676\\
110	49099.8088940385\\
120	53732.5989452952\\
130	58372.0084335881\\
140	63017.2865665018\\
150	67667.8145834364\\
160	72323.0753694149\\
170	76982.6315093594\\
180	81646.1090827612\\
190	86313.1854637061\\
200	90983.579978899\\
};
\addlegendentry{Optimal (Bound) }

\end{axis}

\end{tikzpicture}%
}
  \caption{distances $d_l=10, d_r=30 \text{ km}$}
  \label{fig:rates_2}
\end{subfigure}
\vspace{-5pt}
\begin{subfigure}[b]{0.49\linewidth}
  \centering
%
%
\definecolor{mycolor1}{rgb}{0.14900,0.54900,0.86600}%
\definecolor{mycolor2}{rgb}{0.96000,0.46600,0.16000}%
\definecolor{mycolor3}{rgb}{1.00000,0.90900,0.39200}%
\definecolor{mycolor4}{rgb}{0.85098,0.85098,0.85098}%
\definecolor{mycolor5}{rgb}{0.07059,0.07059,0.07059}%
\definecolor{mycolor6}{rgb}{0.28600,0.85800,0.25000}%

\begin{tikzpicture}

\begin{axis}[%
line width=0.8pt, 
width=1.2in,
height=0.6in,
scale only axis,
scale only axis,
xmin=5,
xmax=30,
xlabel style={font=\color{black}},
xlabel={\# of memories ($N$)},ymin=0,
xlabel style={yshift=0.2cm},
ymax=50,
ylabel style={font=\color{white!15!black},align=center},
ylabel={Rel. diff. \\ to standard rate $[\%]$},
axis background/.style={fill=white},
legend style={legend cell align=left, align=left, draw=white!15!black},
xmajorgrids,
ymajorgrids,
legend style={legend cell align=left, align=left},
legend pos= outer north east
]
\addplot [color=mycolor1, mark options={solid, mycolor1}]
  table[row sep=crcr]{%
2	55.5105518918336\\
4	20.9113576407769\\
6	12.0717798398182\\
8	10.199563138618\\
10	8.62659857380882\\
12	7.32060471849557\\
14	6.21534044676371\\
16	5.78351631377327\\
18	5.84525215252408\\
20	5.36648465286407\\
22	4.89858690457282\\
24	4.68787018127153\\
26	4.27785498613586\\
28	4.19538679479811\\
30	4.06394865344552\\
32	3.79349940335429\\
34	3.79958107757832\\
36	3.67499664814102\\
38	3.66211239760285\\
40	3.50932068971951\\
40	3.46350103047602\\
60	2.76894826936151\\
80	2.46316516627558\\
100	2.17114684883417\\
120	1.93767591853104\\
140	1.86891454870973\\
160	1.63488270678619\\
180	1.56865449567531\\
200	1.54768848141066\\
};

\addplot [color=mycolor2,  mark options={solid, mycolor2}]
  table[row sep=crcr]{%
2	36.201863257275\\
4	30.5687876350993\\
6	29.5949910791049\\
8	28.7055580861297\\
10	28.3758170740999\\
12	28.3452051526615\\
14	28.1849143829248\\
16	28.1767793250622\\
18	28.241156758271\\
20	28.1755138716169\\
22	27.8818629344683\\
24	27.948093809592\\
26	28.2103161590426\\
28	28.2172824438383\\
30	27.9951365392565\\
32	27.9321249923062\\
34	27.9766285441426\\
36	28.0220793152557\\
38	28.0877377073266\\
40	28.0223337450489\\
40	27.9320647326183\\
60	28.0189190294029\\
80	28.012993493429\\
100	28.1234977090467\\
120	28.081307892912\\
140	28.0004164357176\\
160	28.0696978597155\\
180	28.0482454108335\\
200	28.0730965061114\\
};

\addplot [color=mycolor6,  mark options={solid, mycolor6}]
  table[row sep=crcr]{%
2	36.1174996942557\\
4	31.448579078015\\
6	18.9515232760864\\
8	17.4719470732285\\
10	17.6544134081017\\
12	13.869221201051\\
14	13.2835544251244\\
16	12.604470785453\\
18	11.5427955179802\\
20	11.4544146424361\\
22	10.6708195739761\\
24	10.1570235625747\\
26	10.3315448893701\\
28	9.35255672949755\\
30	9.01060308739261\\
32	9.16123222050734\\
34	8.56124068712646\\
36	8.90663503925262\\
38	8.16436254718337\\
40	7.88523332984013\\
40	7.75338513277848\\
60	6.58020935618172\\
80	5.58572364068493\\
100	5.24125516093948\\
120	4.85621985931929\\
140	4.44259736177315\\
160	4.25847388548349\\
180	4.0610062360908\\
200	3.66946556638892\\
};

\end{axis}

\end{tikzpicture}%
  \caption{distances $d_l=20, d_r=30 \text{ km}$}
  \label{fig:rel_diff_1}
\end{subfigure}
\hfill
\begin{subfigure}[b]{0.49\linewidth}
  \centering
   \hspace{15pt}
%
%
\definecolor{mycolor1}{rgb}{0.14900,0.54900,0.86600}%
\definecolor{mycolor2}{rgb}{0.96000,0.46600,0.16000}%
\definecolor{mycolor3}{rgb}{1.00000,0.90900,0.39200}%
\definecolor{mycolor4}{rgb}{0.85098,0.85098,0.85098}%
\definecolor{mycolor5}{rgb}{0.07059,0.07059,0.07059}%
\definecolor{mycolor6}{rgb}{0.28600,0.85800,0.25000}%

\begin{tikzpicture}

\begin{axis}[%
line width=0.8pt, 
width=1.2in,
height=0.6in,
scale only axis,
scale only axis,
xmin=5,
xmax=30,
xlabel style={font=\color{black}},
xlabel={\# of memories ($N$)},ymin=0,
xlabel style={yshift=0.2cm},
ymin=0,
ymax=50,
ylabel style={font=\color{black}},
axis background/.style={fill=white},
legend style={legend cell align=left, align=left, draw=white!15!black},
xmajorgrids,
ymajorgrids,
legend style={legend cell align=left, align=left},
legend pos= outer north east,
yticklabels={}
]
\addplot [color=mycolor1, mark options={solid, mycolor1}]
  table[row sep=crcr]{%
2	67.2629250959392\\
4	19.9994746987398\\
6	15.792127449391\\
8	10.5636611393632\\
10	8.42131113392904\\
12	7.95238173160713\\
14	6.567454018274\\
16	6.39149958645459\\
18	6.1048219749244\\
20	5.58329191559515\\
22	5.11469451291886\\
24	4.88524410579173\\
26	4.67868844102208\\
28	4.53499778274812\\
30	4.42548961137818\\
32	4.26023991498276\\
34	4.03234549894177\\
36	3.95604765813539\\
38	3.8116371112964\\
40	3.66138772280907\\
40	3.67248250885305\\
60	3.03170565188353\\
80	2.6337532352972\\
100	2.29613586634131\\
120	2.13400166291592\\
140	1.94051488121242\\
160	1.85715989165203\\
180	1.7347071934845\\
200	1.55496000291856\\
};

\addplot [color=mycolor2,  mark options={solid, mycolor2}]
  table[row sep=crcr]{%
2	39.978213847296\\
4	38.8916107483155\\
6	38.0754418179233\\
8	37.9593428303803\\
10	37.8592335237564\\
12	37.958647626862\\
14	37.9107778988931\\
16	37.7374860076278\\
18	37.6126680092937\\
20	37.7280486198679\\
22	37.7009167225614\\
24	37.6133052791855\\
26	37.5730235830238\\
28	37.6661655752674\\
30	37.7189414537792\\
32	37.7035948361145\\
34	37.5841782964882\\
36	37.8243574805922\\
38	37.7249860654218\\
40	37.5495203563886\\
40	37.8985472827118\\
60	37.7644772842227\\
80	37.6453020211116\\
100	37.7303845036891\\
120	37.7863066746949\\
140	37.7529865632144\\
160	37.7956224018391\\
180	37.757733810096\\
200	37.7492592792089\\
};

\addplot [color=mycolor6,  mark options={solid, mycolor6}]
  table[row sep=crcr]{%
2	40.2952266516051\\
4	20.417272455093\\
6	20.3185535555055\\
8	14.8028088419348\\
10	16.0925504089396\\
12	12.3463072102986\\
14	12.2821995144397\\
16	10.6975451663961\\
18	10.3555919358708\\
20	11.6579166865554\\
22	9.02851161990547\\
24	9.74450804335394\\
26	8.3346887855941\\
28	8.55059531559656\\
30	9.34852011937486\\
32	7.76413392125976\\
34	8.19154606662054\\
36	7.15765575205115\\
38	7.38363348513201\\
40	8.16003494479907\\
40	8.19955726480994\\
60	6.62957915960994\\
80	5.84078386774771\\
100	5.10868242279639\\
120	4.6113685860365\\
140	4.29854190005504\\
160	3.97090365483087\\
180	3.72291007647456\\
200	3.49264708451421\\
};

\end{axis}

\end{tikzpicture}%
  \caption{distances $d_l=10, d_r=30 \text{ km}$}
  \label{fig:rel_diff_2}
\end{subfigure}
\caption{Asymmetric repeater rate evaluation under different memory allocation regimes: Optimal memory allocation (blue), fixed equal memory allocation (orange), and proportional memory allocation (green). The distances $d_r$ and $d_l$ are the repeater distances to the left and right node, respectively. 
Under optimal allocation, the expected rate is close to the standard repeater rate. 
Under fixed equal allocation, i.e., the left and right bank sizes are both $N/2$, the expected rate is significantly lower with asymmetric distances. Observe the nearly constant relative difference to the standard repeater rate for an arbitrary total number of memories. 
Such a difference improves under proportional allocation, i.e., the right bank size $N_r=\lfloor p_l N /(p_l+p_r)\rfloor$.
}
\label{fig: rates}
\vspace{-10pt}
\end{figure}

\vspace{-5pt}
\section{Numerical Evaluations}
\label{sec: eval}
We numerically evaluate the expected rate and expected fidelity of optimal allocation of memories for asymmetric quantum repeaters compared to different memory allocation regimes.
We set the photon decay rate as $\sigma=0.15 \text{ dB/km}$ and the initial fidelity as $F_0=1$.
Specifically, we evaluate (i) optimal memory allocation according to Lem.~\ref{lemm:optimal_alloc}, (ii) fixed equal memory allocation, i.e., each of the left and right banks is assigned $N/2$ number of memories, (iii) fixed proportional memory allocation by assigning $N_r=\lfloor p_lN/(p_l+p_r)\rfloor$ number of memories to the right bank and the rest to the left bank and (iv) standard repeater. 
Note that the fixed proportional memory allocation is the optimal allocation when the number of unmatched entanglements $\alpha(t)$ is \textit{always zero} (cf. Lem.~\ref{lemm:optimal_alloc}).
However, this does not account for a non-zero number of unmatched entanglements, i.e, $|\alpha(t)|>0$ at any time $t$.
We obtain the stationary expected rate and fidelity using Monte Carlo simulations, which we compare to the rate and fidelity bounds from~\cref{sect:approach} under optimal memory allocation.

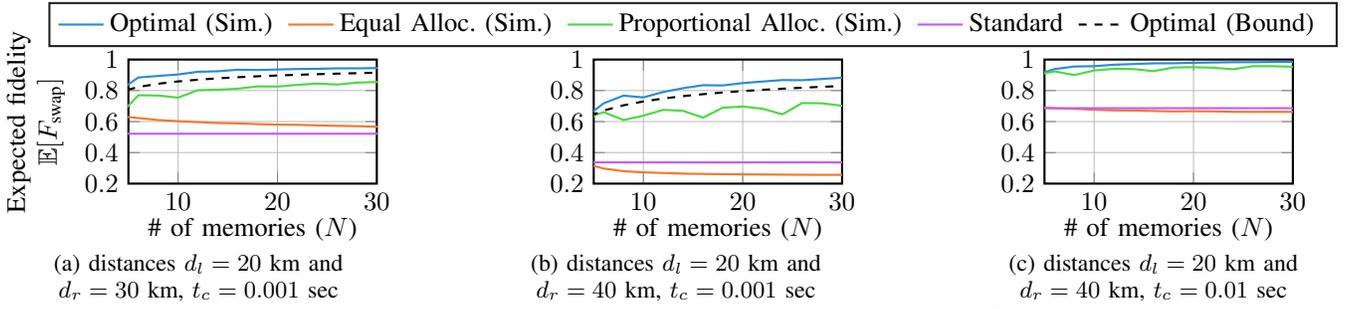
\begin{figure*}[t]
\centering
\begin{subfigure}[b]{0.3\linewidth}
  \centering
%
%
\definecolor{mycolor1}{rgb}{0.14900,0.54900,0.86600}%
\definecolor{mycolor2}{rgb}{0.96000,0.46600,0.16000}%
\definecolor{mycolor3}{rgb}{1.00000,0.90900,0.39200}%
\definecolor{mycolor4}{rgb}{0.75200,0.36000,0.98400}%
\definecolor{mycolor5}{rgb}{0.28600,0.85800,0.25000}%
\definecolor{mycolor6}{rgb}{0.85098,0.85098,0.85098}%
\definecolor{mycolor7}{rgb}{0.07059,0.07059,0.07059}%
\begin{tikzpicture}

\begin{axis}[%
line width=0.8pt,
width=1.3in,
height=0.65in,
scale only axis,
xmin=5,
xmax=30,
xlabel style={font=\color{black}},
xlabel={\# of memories ($N$)},
ymin=0.2,
ymax=1,
ylabel style={font=\color{black}, align=center},
xlabel style={yshift=0.2cm},
ylabel={Expected fidelity \\ $\Ebb[F_{\text{swap}}]$},
axis background/.style={fill=white},
legend style={legend cell align=left, align=left, draw=white!15!black},
xmajorgrids,
ymajorgrids,
legend style={legend cell align=left, align=left},
legend style={at={(0.4,1.1)}, anchor=south, legend columns=5}]
\addplot [color=mycolor1, mark options={solid, mycolor1}]
  table[row sep=crcr]{%
2	0.671011473319746\\
4	0.796320868619204\\
6	0.884466413959583\\
8	0.894044312733422\\
10	0.903836968608089\\
12	0.920976952100639\\
14	0.924315953379366\\
16	0.934159250815363\\
18	0.932586810090418\\
20	0.935762539629861\\
22	0.938801737794964\\
24	0.940004462782271\\
26	0.943243861827343\\
28	0.943223971711183\\
30	0.944928869440214\\
40	0.949819265859161\\
50	0.953470235907975\\
60	0.955987317165895\\
70	0.958352231881682\\
80	0.960351349296406\\
90	0.961998539159481\\
100	0.963572603767229\\
};



\addplot [color=mycolor2,  mark options={solid, mycolor2}]
  table[row sep=crcr]{%
2	0.669495820076281\\
4	0.634945296088999\\
6	0.622831862772225\\
8	0.60998955139005\\
10	0.602388506258005\\
12	0.596485794431063\\
14	0.590593860577862\\
16	0.58818565688349\\
18	0.583611877205377\\
20	0.580072252423838\\
22	0.579358023270613\\
24	0.575383382714274\\
26	0.572354719914295\\
28	0.57055580489157\\
30	0.566358507742145\\
40	0.562513414122044\\
50	0.559451270078307\\
60	0.55676559224577\\
70	0.555529590162634\\
80	0.552539220083262\\
90	0.551941173729169\\
100	0.551957196438823\\
};

\addplot [color=mycolor5,  mark options={solid, mycolor5}]
  table[row sep=crcr]{%
2	0.670058667064872\\
4	0.635976156889165\\
6	0.769543731220128\\
8	0.768024021098216\\
10	0.754394282089642\\
12	0.802544827328302\\
14	0.805213244029005\\
16	0.811858137658992\\
18	0.826768894259372\\
20	0.825639758522655\\
22	0.836682648121455\\
24	0.845139747407298\\
26	0.838760363517511\\
28	0.851229992607327\\
30	0.855071063146851\\
40	0.872963995334687\\
50	0.882589109432383\\
60	0.892559778752255\\
70	0.898724549222711\\
80	0.903272707261291\\
90	0.909228283852352\\
100	0.911429246399672\\
};
\addplot [color=mycolor4, mark options={solid, mycolor4}]
  table[row sep=crcr]{%
2	0.521583\\
4	0.521583\\
6	0.521583\\
8	0.521583\\
10	0.521583\\
12	0.521583\\
14	0.521583\\
16	0.521583\\
18	0.521583\\
20	0.521583\\
22	0.521583\\
24	0.521583\\
26	0.521583\\
28	0.521583\\
30	0.521583\\
40	0.521583\\
50	0.521583\\
60	0.521583\\
70	0.521583\\
80	0.521583\\
90	0.521583\\
100	0.521583\\
};

\addplot [color=black,dashed, mark options={solid, black}]
  table[row sep=crcr]{%
2	0.720479314900594\\
4	0.789333606222139\\
6	0.822975253936735\\
8	0.844019769178977\\
10	0.858823265042552\\
12	0.869986912347516\\
14	0.878804473337182\\
16	0.886003305546917\\
18	0.892028568773278\\
20	0.897170310701262\\
22	0.901626842564002\\
24	0.905539045711658\\
26	0.90901017370397\\
28	0.912117888993937\\
30	0.914921898028094\\
40	0.925734747354251\\
50	0.9332145670354\\
60	0.938789005253957\\
70	0.943152848280132\\
80	0.946690557583456\\
90	0.949634562034649\\
100	0.952134813920081\\
};


\end{axis}

\end{tikzpicture}%
   \caption{distances $d_l=20 \text{ km}$ and $d_r=30 \text{ km}$, $t_c=0.001 \text{ sec}$}
  \label{fig:fid_1}
\end{subfigure}
\hfill
\begin{subfigure}[b]{0.3\linewidth}
\hspace{-16em}
%
%
%
\definecolor{mycolor1}{rgb}{0.14900,0.54900,0.86600}%
\definecolor{mycolor2}{rgb}{0.96000,0.46600,0.16000}%
\definecolor{mycolor3}{rgb}{1.00000,0.90900,0.39200}%
\definecolor{mycolor4}{rgb}{0.75200,0.36000,0.98400}%
\definecolor{mycolor5}{rgb}{0.28600,0.85800,0.25000}%
\definecolor{mycolor6}{rgb}{0.85098,0.85098,0.85098}%
\definecolor{mycolor7}{rgb}{0.07059,0.07059,0.07059}%
\begin{tikzpicture}

\begin{axis}[%
line width=0.8pt,
width=1.3in,
height=0.65in,
scale only axis,
xmin=5,
xmax=30,
xlabel style={font=\color{black}},
xlabel={\# of memories ($N$)},
ymin=0.2,
ymax=1,
ylabel style={font=\color{black}},
xlabel style={yshift=0.2cm},
axis background/.style={fill=white},
legend style={legend cell align=left, align=left, draw=white!15!black},
xmajorgrids,
ymajorgrids,
legend style={legend cell align=left, align=left},
legend style={at={(0.4,1.1)}, anchor=south, legend columns=5}]
\addplot [color=mycolor1, mark options={solid, mycolor1}]
  table[row sep=crcr]{%
2	0.416906186919465\\
4	0.618271515336997\\
6	0.717835619650823\\
8	0.767193571091479\\
10	0.755936530892898\\
12	0.790657154283724\\
14	0.815111340294563\\
16	0.834520096205667\\
18	0.833008209521672\\
20	0.848286041495776\\
22	0.858193834088175\\
24	0.867647351075494\\
26	0.866783621465581\\
28	0.874771958519697\\
30	0.882813614281321\\
40	0.898835534876835\\
50	0.906990512259669\\
60	0.914068031751786\\
70	0.919115358980103\\
80	0.922433758858753\\
90	0.926535623387991\\
100	0.929028970210521\\
};
\addlegendentry{Optimal (Sim.) }


\addplot [color=mycolor2, mark options={solid, mycolor2}]
  table[row sep=crcr]{%
2	0.42461797465389\\
4	0.332125482201625\\
6	0.297166144347971\\
8	0.279962574283291\\
10	0.272808900812282\\
12	0.268560867696506\\
14	0.264147157303898\\
16	0.26254719285735\\
18	0.260945233772085\\
20	0.259387806137365\\
22	0.258648199451164\\
24	0.25786419785096\\
26	0.257225627370346\\
28	0.256600283734341\\
30	0.256400250179929\\
40	0.25496493032138\\
50	0.254376146785948\\
60	0.254004227040107\\
70	0.253665258101641\\
80	0.253294528573556\\
90	0.253325589244582\\
100	0.253216807235501\\
};
\addlegendentry{Equal Alloc. (Sim.) }
\addplot [color=mycolor5,  mark options={solid, mycolor5}]
  table[row sep=crcr]{%
2	0.409936354555387\\
4	0.615100067876735\\
6	0.661128866869413\\
8	0.60992478325451\\
10	0.637776310166229\\
12	0.675632288291575\\
14	0.66911628045467\\
16	0.625372616452457\\
18	0.689452666174467\\
20	0.697063382384474\\
22	0.682581743654204\\
24	0.646749174102457\\
26	0.719784026532611\\
28	0.717859257769426\\
30	0.702032784318573\\
40	0.688622656776996\\
50	0.765414953853723\\
60	0.764890987310127\\
70	0.761796512437988\\
80	0.757176393061229\\
90	0.750644884764336\\
100	0.800320382838626\\
};
\addlegendentry{Proportional Alloc. (Sim.) }

\addplot [color=mycolor4,  mark options={solid, mycolor4}]
  table[row sep=crcr]{%
2	0.336792\\
4	0.336792\\
6	0.336792\\
8	0.336792\\
10	0.336792\\
12	0.336792\\
14	0.336792\\
16	0.336792\\
18	0.336792\\
20	0.336792\\
22	0.336792\\
24	0.336792\\
26	0.336792\\
28	0.336792\\
30	0.336792\\
40	0.336792\\
50	0.336792\\
60	0.336792\\
70	0.336792\\
80	0.336792\\
90	0.336792\\
100	0.336792\\
};
\addlegendentry{Standard }
\addplot [color=black,dashed, mark options={solid, black}]
  table[row sep=crcr]{%
2	0.5259894438003\\
4	0.619877204223036\\
6	0.671108293181503\\
8	0.704963825869953\\
10	0.729617561628342\\
12	0.74866925868843\\
14	0.763997355618788\\
16	0.776695265943484\\
18	0.787450328505456\\
20	0.796720094064707\\
22	0.80482298823626\\
24	0.811988629676907\\
26	0.81838748423268\\
28	0.82414921159722\\
30	0.829374482127982\\
40	0.84976092838835\\
50	0.86408360833364\\
60	0.874875143723813\\
70	0.883393121900963\\
80	0.890343712820334\\
90	0.896158707548104\\
100	0.901119233124725\\
};
\addlegendentry{Optimal (Bound) }


\end{axis}

\end{tikzpicture}%
  \caption{distances $d_l=20 \text{ km}$ and $d_r=40 \text{ km}$, $t_c=0.001 \text{ sec}$}
  \label{fig:fid_2}
\end{subfigure}
\hfill
\begin{subfigure}[b]{0.3\linewidth}
  \centering
%
%
\definecolor{mycolor1}{rgb}{0.14900,0.54900,0.86600}%
\definecolor{mycolor2}{rgb}{0.96000,0.46600,0.16000}%
\definecolor{mycolor3}{rgb}{1.00000,0.90900,0.39200}%
\definecolor{mycolor4}{rgb}{0.75200,0.36000,0.98400}%
\definecolor{mycolor5}{rgb}{0.28600,0.85800,0.25000}%
\definecolor{mycolor6}{rgb}{0.42300,0.95600,1.00000}%
\definecolor{mycolor7}{rgb}{0.94900,0.40300,0.77200}%
\definecolor{mycolor8}{rgb}{0.85098,0.85098,0.85098}%
\definecolor{mycolor9}{rgb}{0.07059,0.07059,0.07059}%
\begin{tikzpicture}

\begin{axis}[%
line width=0.8pt,
width=1.3in,
height=0.65in,
scale only axis,
xmin=5,
xmax=30,
xlabel style={font=\color{black}},
xlabel={\# of memories ($N$)},
ymin=0.2,
ymax=1,
ylabel style={font=\color{black}},
xlabel style={yshift=0.2cm},
axis background/.style={fill=white},
legend style={legend cell align=left, align=left, draw=white!15!black},
xmajorgrids,
ymajorgrids,
legend style={legend cell align=left, align=left},
legend pos = outer north east,
legend style={at={(0.5,1.1)}, anchor=south, legend columns=1},
]
\addplot [color=mycolor1, mark options={solid, mycolor1}]  table[row sep=crcr]{%
2	0.721977158458368\\
4	0.898282562045745\\
6	0.938476293123214\\
8	0.954981269572006\\
10	0.957047924141019\\
12	0.966477139151303\\
14	0.970945135524784\\
16	0.975455131716603\\
18	0.975932039193233\\
20	0.979082052137723\\
22	0.980647708569055\\
24	0.982675376008593\\
26	0.982829011367412\\
28	0.984122182009935\\
30	0.985062661956431\\
40	0.987791746398477\\
50	0.98905712998348\\
60	0.990059834629752\\
70	0.99086548297128\\
80	0.991326863695953\\
90	0.991814251436924\\
100	0.992173957386919\\
};



\addplot [color=mycolor2,  mark options={solid, mycolor2}]
  table[row sep=crcr]{%
2	0.72492168237882\\
4	0.695715599455132\\
6	0.686162545440172\\
8	0.682349452370671\\
10	0.676174417673103\\
12	0.672182101254341\\
14	0.671127632556138\\
16	0.668218980026639\\
18	0.665532818651787\\
20	0.667166043175413\\
22	0.665768332412536\\
24	0.663009887677478\\
26	0.66257393360822\\
28	0.662791855867504\\
30	0.662393819994509\\
40	0.658650080050864\\
50	0.660634131887341\\
60	0.659512412980438\\
70	0.658747858444522\\
80	0.657633023210653\\
90	0.657899094165345\\
100	0.658915978734419\\
};

\addplot [color=mycolor5, mark options={solid, mycolor5}]
table[row sep=crcr]{%
2	0.728078903822506\\
4	0.895996662983025\\
6	0.923745068777874\\
8	0.899821193679209\\
10	0.929359800646362\\
12	0.940237483149165\\
14	0.938284942953769\\
16	0.925005964793839\\
18	0.948536166900482\\
20	0.950722415191336\\
22	0.947215931082554\\
24	0.93755489277452\\
26	0.957211976444494\\
28	0.956821129307598\\
30	0.95337063074952\\
40	0.951575777644966\\
50	0.967086592793796\\
60	0.967745649367006\\
70	0.96760853820518\\
80	0.96719997689346\\
90	0.963678523763903\\
100	0.974971513911742\\
};

\addplot [color=mycolor4,  mark options={solid, mycolor4}]
  table[row sep=crcr]{%
2	0.686273\\
4	0.686273\\
6	0.686273\\
8	0.686273\\
10	0.686273\\
12	0.686273\\
14	0.686273\\
16	0.686273\\
18	0.686273\\
20	0.686273\\
22	0.686273\\
24	0.686273\\
26	0.686273\\
28	0.686273\\
30	0.686273\\
40	0.686273\\
50	0.686273\\
60	0.686273\\
70	0.686273\\
80	0.686273\\
90	0.686273\\
100	0.686273\\
};




\end{axis}

\end{tikzpicture}%
  \caption{distances $d_l=20 \text{ km}$ and $d_r=40 \text{ km}$, $t_c=0.01 \text{ sec}$}
  \label{fig:fid_3}
\end{subfigure}
   \captionsetup{skip=0pt} 
\caption{Asymmetric repeater fidelity evaluation under different memory allocation regimes: Optimal memory allocation (blue), fixed equal memory allocation (orange), and proportional memory allocation (green). The figure compares the expected swapped fidelity considering different distances and coherence times $t_c$.
Optimal allocation is superior while the fidelity of the standard repeater is significantly lower.  
}
\label{fig:fid}
\vspace{-20pt}
\end{figure*}

\vspace{-5pt}
\subsection{Rate Evaluation}
First, we evaluate the expected rate in~\cref{fig: rates}
for increasing total number of memories $N$ and considering different repeater-node distances. We fix the right repeater-node distance to $d_r=30\text{ km}$ and the left repeater-node distances to $d_l=10\text{ km}$ or $d_l=20\text{ km}$ in~\cref{fig:rates_1} or~\cref{fig:rates_2}, respectively.
We show in~\cref{fig:rates_1,fig:rates_2} that the expected rate under optimal allocation is very close to the standard repeater rate.
Observe that the rate bound under optimal allocation is tight. 
Figure~\ref{fig:rel_diff_1} shows the relative rate difference to the standard rate, i.e., $(R_{\text{std}}-R)/R_{\text{std}}$, which highlights the constant relative difference of the fixed equal allocation when the distance $d_l=20\text{ km}$.
Hence, under fixed equal allocation, not accounting for asymmetric links results in a significantly lower rate, shown by the high relative error (nearly constant) in~\cref{fig:rel_diff_1}. 
Moreover, for $d_l=10\text{ km}$,~\cref{fig:rel_diff_2} shows that the relative difference becomes larger for a higher ratio $d_r/d_l$.
This demonstrates the necessity of optimal dynamic allocation.

We observe that fixed memory allocation proportional to the repeater-node distances further improves the rate, as show in~\cref{fig:rel_diff_1,fig:rel_diff_2}.
As proportional allocation is mismatch agnostic, it achieves a worse rate than optimal allocation, where the gap in the relative error is shown in~\cref{fig:rel_diff_1,fig:rel_diff_2} to be nearly constant for a significant number of memories.
Notably, larger errors under optimal allocation for a smaller total number of memories stem from possible violations of the optimal condition as discussed in~\cref{sec:Alloc_viol}.

\label{sec:numerical_validation}

\vspace{-5pt}
\subsection{Fidelity Evaluation}
Next, we evaluate in~\cref{fig:fid} the expected fidelity of end-to-end entanglements, i.e., the fidelity $\Ebb[F_{\text{swap}}]$ considering different repeater-node distances, where we set the left repeater-node distance as $d_l=20\text{ km}$ and the right repeater-node distances as $d_r=30\text{ km}$ in~\cref{fig:fid_1} or $d_r=40\text{ km}$ in~\cref{fig:fid_2,fig:fid_3}. 
Additionally, we increase the decoherence time from $t_c=0.001$ in~\cref{fig:fid_1,fig:fid_2} to $t_c=0.01$ in~\cref{fig:fid_3}.

Figure~\ref{fig:fid} shows that optimal allocation is superior to the standard repeater and the equal and proportional allocations since it reduces the number of unmatched entanglements, thereby reducing the fidelity decay.
Observe that the fidelity bound under optimal allocation from~\cref{sec:fid_LB} is tight.
In contrast, \textit{the sequential entanglement generation in a standard repeater} yields significantly low fidelity as the fidelity of the first-formed entanglement decays  while waiting for the entanglement generation with the opposite side.
\textit{The possible matching between any two memories in our considered repeater significantly reduces such waiting times}. 
Observe that fixed proportional memory allocation yields significantly higher fidelity than the standard repeater. 

Figures~\ref{fig:fid_1} and \ref{fig:fid_2} highlight that the fidelity difference between optimal allocation and proportional allocation is more significant compared to the rate difference in~\cref{fig:rates_1,fig:rates_2}.
The reason is that the fidelity decays exponentially with the mismatch, precisely, the expected fidelity decays as $\Ebb[F_\text{swap}] \sim \e^{-\Ebb[\abs{\alpha}] \tau_{\text{round}}/ t_c}$ (see~(\ref{eq:F_LB}-\ref{eq:F_LB_1})).
Accordingly, the effect of the unmatched entanglements is sensitive to the ratio between the repeater round time $\tau_\text{round}$ and the coherence time $t_c$.
Such sensitivity is shown by comparing the fidelity differences.
First, by comparing~\cref{fig:fid_1,fig:fid_2} where the distance of the repeater increases from $d_r=30\text{ km}$ to $d_r=40\text{ km}$, the difference in fidelity between the optimal allocation and proportional allocation is higher.
Recall that the repeater round time from~\cref{eq:round_time} depends on repeater-node distances.
Second, for an order-of-magnitude longer coherence time in~\cref{fig:fid_3} ($t_c\!=\!0.01\text{ sec})$ than in~\cref{fig:fid_2} ($t_c\!=\!0.001\text{ sec})$, the difference in fidelity between optimal allocation and proportional allocation is smaller.
Notably, a common behavior in~\cref{fig:fid} is that equal memory allocation is detrimental to the fidelity, which may result in even worse fidelity than the standard repeater.

\vspace{-5pt}
\subsection{Hard-Cutoff regime evaluation}
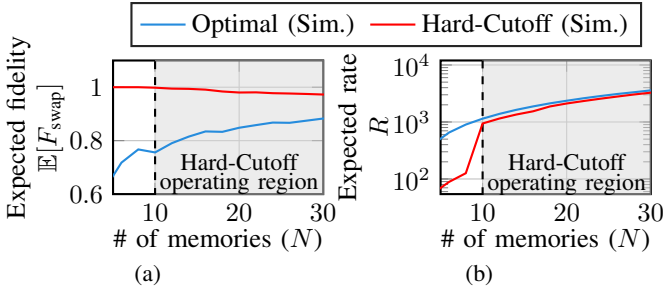
\begin{figure}[t]
\centering
\hspace{-2em}
\begin{subfigure}[b]{0.45\linewidth}
  \centering
%
%
%
\definecolor{mycolor1}{rgb}{0.14900,0.54900,0.86600}%
\definecolor{mycolor2}{rgb}{0.96000,0.46600,0.16000}%
\definecolor{mycolor3}{rgb}{1.00000,0.90900,0.39200}%
\definecolor{mycolor4}{rgb}{0.75200,0.36000,0.98400}%
\definecolor{mycolor5}{rgb}{0.28600,0.85800,0.25000}%
\definecolor{mycolor6}{rgb}{0.85098,0.85098,0.85098}%
\definecolor{mycolor7}{rgb}{0.07059,0.07059,0.07059}%
\begin{tikzpicture}

\begin{axis}[%
line width=0.8pt,
width=1.1in,
height=0.7in,
scale only axis,
xmin=5,
xmax=30,
xlabel style={font=\color{black}},
ylabel={Expected fidelity \\ $\Ebb[F_{\text{swap}}]$},
ymin=0.6,
ymax=1.1,
ylabel style={font=\color{black},align=center},
xlabel={\# of memories ($N$)},
ylabel style={yshift=-0.2cm},
xlabel style={yshift=0.2cm},
axis background/.style={fill=white},
legend style={legend cell align=left, align=left, draw=white!15!black},
xmajorgrids,
ymajorgrids,
legend style={legend cell align=left, align=left},
legend style={at={(1.3,1.1)}, anchor=south, legend columns=8}]
\draw[fill=gray!30, draw=none, opacity=0.5]
  (axis cs:10,0) rectangle (axis cs:100,2);
\draw[dashed] (axis cs:10,0) -- (axis cs:10, 2) node[right]{};
\node at (axis cs:20,0.72){\small Hard-Cutoff};
\node at (axis cs:20,0.64){\small operating region};
\addplot [color=mycolor1, mark options={solid, mycolor1}]
  table[row sep=crcr]{%
2	0.416906186919465\\
4	0.618271515336997\\
6	0.717835619650823\\
8	0.767193571091479\\
10	0.755936530892898\\
12	0.790657154283724\\
14	0.815111340294563\\
16	0.834520096205667\\
18	0.833008209521672\\
20	0.848286041495776\\
22	0.858193834088175\\
24	0.867647351075494\\
26	0.866783621465581\\
28	0.874771958519697\\
30	0.882813614281321\\
40	0.898835534876835\\
50	0.906990512259669\\
60	0.914068031751786\\
70	0.919115358980103\\
80	0.922433758858753\\
90	0.926535623387991\\
100	0.929028970210521\\
};
\addlegendentry{Optimal (Sim.) }
\addplot [color=red,  mark options={solid, red}]
  table[row sep=crcr]{%
2	1\\
4	1\\
6	1\\
8	1\\
10	0.998163213223039\\
12	0.994730453406013\\
14	0.993830902625925\\
16	0.990941979912433\\
18	0.984048995369721\\
20	0.980175176507918\\
22	0.980771916719801\\
24	0.97764995875469\\
26	0.976444886089705\\
28	0.974738203188951\\
30	0.972648598485706\\
40	0.969901056576353\\
50	0.96970199443552\\
60	0.937765308615694\\
70	0.941719469509401\\
80	0.945782053680493\\
90	0.947495229205245\\
100	0.949460937895697\\
};
\addlegendentry{Hard-Cutoff (Sim.)}

\end{axis}

\end{tikzpicture}%
  \caption{}
  \label{fig:hard-cutoff_1}
\end{subfigure}
 \hspace{0.5em}
 \begin{subfigure}[b]{0.45\linewidth}
%
%
\definecolor{mycolor1}{rgb}{0.14900,0.54900,0.86600}%
\definecolor{mycolor2}{rgb}{0.96000,0.46600,0.16000}%
\definecolor{mycolor3}{rgb}{1.00000,0.90900,0.39200}%
\definecolor{mycolor4}{rgb}{0.75200,0.36000,0.98400}%
\definecolor{mycolor5}{rgb}{0.28600,0.85800,0.25000}%
\definecolor{mycolor6}{rgb}{0.42300,0.95600,1.00000}%
\definecolor{mycolor7}{rgb}{0.94900,0.40300,0.77200}%
\definecolor{mycolor8}{rgb}{0.85098,0.85098,0.85098}%
\definecolor{mycolor9}{rgb}{0.07059,0.07059,0.07059}%
\begin{tikzpicture}

\begin{axis}[%
line width=0.8pt,
ymode=log,
width=1.1in,
height=0.7in,
scale only axis,
scale only axis,
xmin=5,
xmax=30,
ylabel style={yshift=-0.2cm},
xlabel style={yshift=0.2cm},
xlabel style={font=\color{black}},
xlabel={\# of memories ($N$)},
ymin=0,
ymax=12000,
ylabel style={font=\color{black}, align=center},
ylabel={Expected rate \\ $R$},
axis background/.style={fill=white},
legend style={legend cell align=left, align=left, draw=white!15!black},
xmajorgrids,
ymajorgrids,
legend style={legend cell align=left, align=left},
legend pos = outer north east,
legend style={at={(0.5,1.1)}, anchor=south, legend columns=1},
]
\draw[fill=gray!30, draw=none, opacity=0.5]
  (axis cs:10,0) rectangle (axis cs:100,100000);

\draw[dashed] (axis cs:10,12000) -- (axis cs:10, 10) node[right]{};
\node at (axis cs:20,200){\small Hard-Cutoff}; \\ 
\node at (axis cs:20,80){\small operating region};
\addplot [color=mycolor1,  mark options={solid, mycolor1}]
  table[row sep=crcr]{%
2	136.1625\\
4	397.125\\
6	652.35\\
8	901.05\\
10	1143.3375\\
12	1387.9125\\
14	1633.1625\\
16	1875.375\\
18	2121.15\\
20	2358.4125\\
22	2606.7\\
24	2848.5375\\
26	3095.625\\
28	3325.7625\\
30	3583.275\\
40	4823.85\\
50	6028.1625\\
60	7243.3125\\
70	8447.2125\\
80	9670.8\\
90	10891.05\\
100	12111.3375\\
};

\addplot [color=red,mark options={solid, red}]
  table[row sep=crcr]{%
2	19.35\\
4	54.6\\
6	88.65\\
8	126.1875\\
10	940.9875\\
12	1150.05\\
14	1350.4125\\
16	1539.1125\\
18	1871.85\\
20	2110.875\\
22	2329.575\\
24	2559.15\\
26	2791.5375\\
28	3034.65\\
30	3268.2\\
40	4456.95\\
50	5591.925\\
60	7120.275\\
70	8321.8125\\
80	9513.45\\
90	10716.45\\
100	11892.2625\\
};

\end{axis}

\end{tikzpicture}%
  \caption{}
  \label{fig:hard-cutoff_2}
\end{subfigure}
   \vspace{-5pt}
\caption{Qualitative example of an asymmetric repeater evaluation of (a) the expected end-to-end fidelity and (b) the expected rate under the memory allocations: (i) Pure optimal memory allocation, (ii) Hard-Cutoff optimal allocation, which drops excess unmatched entanglements.
The repeater node distances $d_l=20 \text{ km}$ and $d_r=40 \text{ km}$ and coherence time $t_c=0.001$.
The figure highlights two regions: (i) Hard-cutoff operating region (gray) and (ii) trade-off region (white). The cutoff value $N=10$ is fixed depending on the example parameters. 
In the hard-cutoff operating region, we observe that hard-cutoff allocation is superior as it  significantly improves the fidelity in (a) while sacrificing a negligible rate in (b). 
In the trade-off region, both memory allocations provide a rate-fidelity trade-off.
}
\label{fig:hard-cutoff}
\vspace{-15pt}
\end{figure}

In~\cref{fig:hard-cutoff}, we evaluate the empirical hard-cutoff allocation regime from~\cref{sec:Alloc_viol} compared to pure optimal allocation from Lem.~\ref{lemm:optimal_alloc}.  
We set the repeater node distances as $d_l=20 \text{ km}$ and $d_r=40 \text{ km}$ and coherence time as $t_c=0.001$.
Recall that the hard-cutoff regime drops excess unmatched entanglements exceeding the threshold in~\cref{eq:viol_1}.
As expected,~\cref{fig:hard-cutoff} shows that the hard-cutoff regime has better fidelity and lower rates for different numbers of memories.
For the given parameterization, we identify two different regions: (i) hard-cutoff operating region for the total number of memories $N\!\!\geq\!\!10$ and (ii) trade-off region when $N\!\!<\!\!10$. 
Remarkably,~\cref{fig:hard-cutoff} shows for a number of memories exceeding $N\!\!=\!\!10$ that the fidelity of the hard-cutoff regime is significantly better with nearly equal rate compared to pure optimal allocation.  
In that region, using hard-cutoff is  favorable. 
For a smaller number of memories ($N<10$), a rate-fidelity trade-off is achieved. 
Specifically, for a few memories, the violation threshold is almost zero (cf.~\cref{eq:viol_1}).
As a result, the hard-cutoff regime nearly resets all the memories, thus yielding high fidelity at the expense of an order of magnitude lower rate.
As the number of memories increases, the violation threshold linearly increases with $N$ (cf.~\cref{eq:viol_1}) and the probability of larger unmatched entanglements decreases exponentially with $N$.
This results in the hard-cutoff regime advantage, which diminishes for asymptotically large number of memories.

\section{Application to multi-repeater systems: Insights and outlook}
\begin{figure}[t]
    \centering
    \begin{subfigure}[b]{0.99\linewidth}
        \centering
        \includegraphics[width=0.9\linewidth, trim={0 0 0 1.5em},clip]{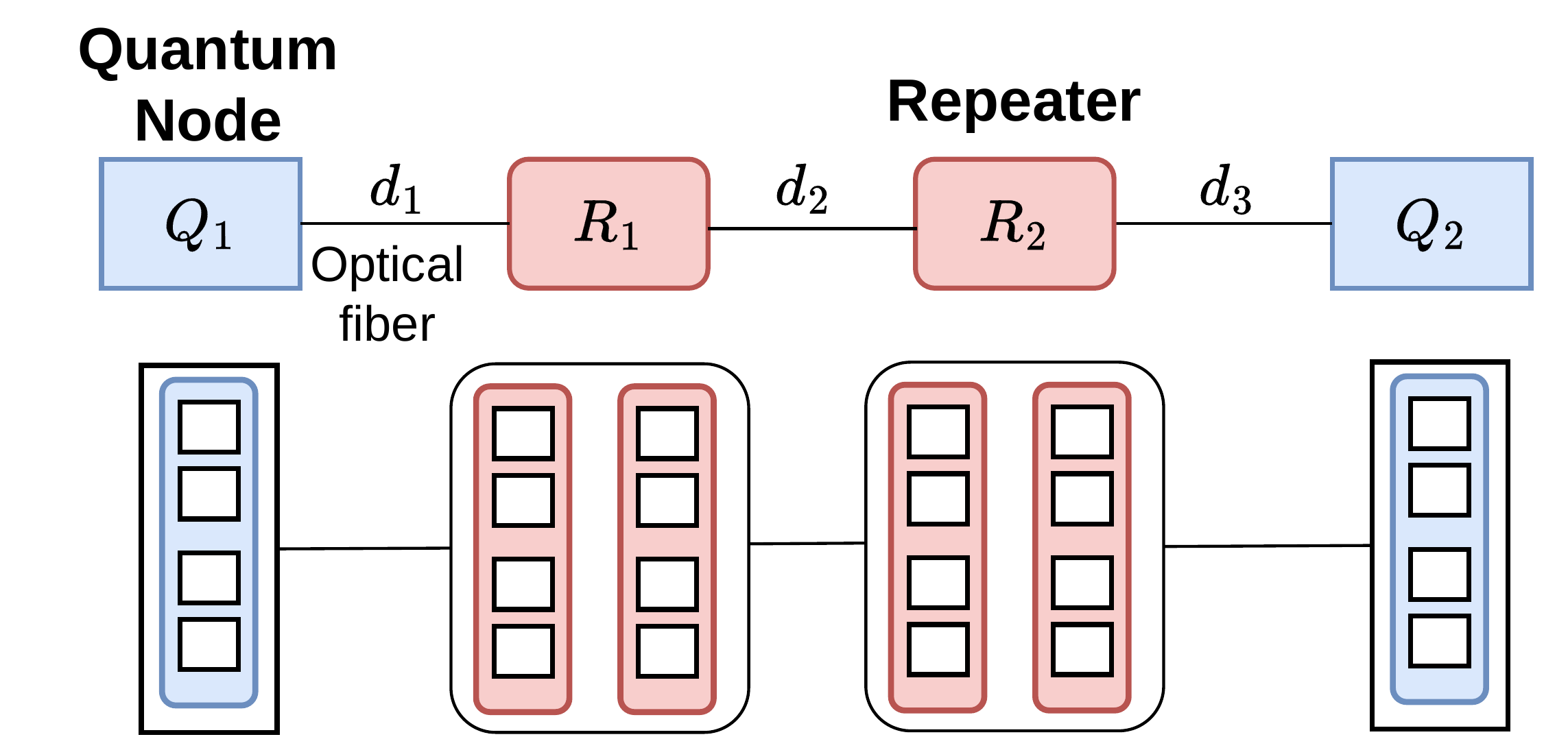}
        \caption{}
        \label{fig:multi_hop_model}
    \end{subfigure}
    \\
\vspace{5pt}
    \hspace{-2em}
    \begin{subfigure}{1\linewidth}
        \begin{subfigure}{0.45\linewidth}
%
%
\definecolor{mycolor1}{rgb}{0.14900,0.54900,0.86600}%
\definecolor{mycolor2}{rgb}{0.96000,0.46600,0.16000}%
\definecolor{mycolor3}{rgb}{1.00000,0.90900,0.39200}%
\definecolor{mycolor4}{rgb}{0.75200,0.36000,0.98400}%
\definecolor{mycolor5}{rgb}{0.28600,0.85800,0.25000}%
\definecolor{mycolor6}{rgb}{0.85098,0.85098,0.85098}%
\definecolor{mycolor7}{rgb}{0.07059,0.07059,0.07059}%
\begin{tikzpicture}

\begin{axis}[%
line width=0.8pt,
width=1.1in,
height=0.7in,
scale only axis,
xmin=10,
xmax=30,
xlabel style={font=\color{black}},
xlabel={\# of memories ($N$)},
ymin=0.2,
ymax=1,
ylabel style={font=\color{black}, align=center},
xlabel style={yshift=0.2cm},
ylabel style={yshift=-0.2cm},
ylabel={Expected fidelity \\ $\Ebb[F_{\text{swap}}]$},
axis background/.style={fill=white},
legend style={legend cell align=left, align=left, draw=white!15!black},
xmajorgrids,
ymajorgrids,
legend style={legend cell align=left, align=left},
legend style={at={(1.1,1.2)}, anchor=south, legend columns=5}
]

\addplot [color=mycolor1, line width=1.2pt,mark options={solid, mycolor1}]
  table[row sep=crcr]{%
2	0.394002716214562\\
4	0.579183575726211\\
6	0.668812330826157\\
8	0.712402450563204\\
10	0.73932673797565\\
12	0.765919493463987\\
14	0.784146073079402\\
16	0.801885427585386\\
18	0.808456393876629\\
20	0.818925912856535\\
22	0.82610618585981\\
24	0.833780798182674\\
26	0.839425245503768\\
28	0.845120492986253\\
30	0.851227650904059\\
40	0.870301222881013\\
50	0.883967830722078\\
60	0.893872761439473\\
70	0.901481057180945\\
80	0.9073375583655\\
90	0.91246526692723\\
100	0.916969580268204\\
110	0.920762273281238\\
120	0.92415232619021\\
130	0.926878594173482\\
140	0.929684103335513\\
150	0.93181709479104\\
160	0.934125167843051\\
170	0.935946487111994\\
180	0.937561237767809\\
190	0.939112981774403\\
200	0.940864791489619\\
};
\addlegendentry{Optimal (Sim.)}

\addplot [color=mycolor2,line width=1.2pt, mark options={solid, mycolor2}]
  table[row sep=crcr]{%
2	0.412399089348473\\
4	0.439743790486549\\
6	0.445794971322756\\
8	0.445993550323697\\
10	0.441660746558095\\
12	0.433354950382463\\
14	0.431325559171188\\
16	0.427732768239078\\
18	0.421925451278934\\
20	0.417441986304134\\
22	0.415220258008797\\
24	0.408970897125925\\
26	0.408784118680166\\
28	0.404970217844182\\
30	0.403381509003248\\
40	0.392724549174605\\
50	0.386211808245557\\
60	0.383651928954847\\
70	0.382045187883165\\
80	0.378110107303892\\
90	0.377081881212671\\
100	0.375986023631361\\
110	0.375563268041582\\
120	0.37423082777168\\
130	0.374163830018116\\
140	0.373094530916592\\
150	0.371857575149382\\
160	0.372482229407897\\
170	0.370794170620388\\
180	0.370464521732361\\
190	0.371688480213801\\
200	0.371232022699601\\
};
\addlegendentry{Equal Alloc. (Sim.)}

\addplot [color=mycolor4, mark options={solid, mycolor4}]
  table[row sep=crcr]{%
2	0.348343\\
4	0.348343\\
6	0.348343\\
8	0.348343\\
10	0.348343\\
12	0.348343\\
14	0.348343\\
16	0.348343\\
18	0.348343\\
20	0.348343\\
22	0.348343\\
24	0.348343\\
26	0.348343\\
28	0.348343\\
30	0.348343\\
40	0.348343\\
50	0.348343\\
60	0.348343\\
70	0.348343\\
80	0.348343\\
90	0.348343\\
100	0.348343\\
};
\addlegendentry{Standard}

\end{axis}

\end{tikzpicture}%
        \end{subfigure}
        \hspace{10pt}
        \begin{subfigure}[b]{0.45\linewidth}
%
%
\definecolor{mycolor1}{rgb}{0.14900,0.54900,0.86600}%
\definecolor{mycolor2}{rgb}{0.96000,0.46600,0.16000}%
\definecolor{mycolor3}{rgb}{1.00000,0.90900,0.39200}%
\definecolor{mycolor4}{rgb}{0.75200,0.36000,0.98400}%
\definecolor{mycolor5}{rgb}{0.85098,0.85098,0.85098}%
\definecolor{mycolor6}{rgb}{0.07059,0.07059,0.07059}%
\begin{tikzpicture}

\begin{axis}[%
line width=0.8pt, 
ymode=log,
width=1.1in,
height=0.7in,
scale only axis,
xmin=10,
xmax=30,
xlabel style={font=\color{black}},
xlabel={\# of memories ($N$)},
xlabel style={yshift=0.2cm},
ymin=500,
ymax=50000,
ylabel style={yshift=-0.2cm},
ylabel style={font=\color{black},align=center},
ylabel={Expected rate\\$R$},
axis background/.style={fill=white},
legend style={legend cell align=left, align=left, draw=white!15!black,at={(0.5,1.1)}, anchor=south, font=\small},
xmajorgrids,
ymajorgrids
]
\addplot [color=mycolor1,mark options={solid, mycolor1}, forget plot]
  table[row sep=crcr]{%
2	399.1\\
4	1222.2\\
6	1975.6\\
8	2654.9\\
10	3378.9\\
12	4139.5\\
14	4838.8\\
16	5630.1\\
18	6301.4\\
20	7078.5\\
22	7852.8\\
24	8568.2\\
26	9379.4\\
28	10095.3\\
30	10921.1\\
40	14729.8\\
50	18599\\
60	22544.2\\
70	26445.4\\
80	30297.4\\
90	34387\\
100	38348\\
110	42335.7\\
120	46277\\
130	50278.9\\
140	54305.3\\
150	58344.2\\
160	62358.6\\
170	66314.8\\
180	70353.3\\
190	74387.7\\
200	78453.2\\
};

\addplot [color=mycolor2,mark options={solid, mycolor2},forget plot]
  table[row sep=crcr]{%
2	492\\
4	1103.5\\
6	1727.6\\
8	2337.3\\
10	2952.7\\
12	3542.7\\
14	4171\\
16	4791.8\\
18	5368.2\\
20	5957.3\\
22	6589.8\\
24	7156.2\\
26	7785.3\\
28	8403.2\\
30	8989.9\\
40	11933.3\\
50	14867.1\\
60	17877.2\\
70	20917.4\\
80	23834.6\\
90	26848.8\\
100	29804.7\\
110	32869.9\\
120	35793.9\\
130	38863.7\\
140	41781.5\\
150	44715.1\\
160	47819.9\\
170	50647.6\\
180	53645.3\\
190	56786.1\\
200	59782.6\\
};


\addplot [color=mycolor4, forget plot]
  table[row sep=crcr]{%
2	835.880000000000\\
4	1671.77000000000\\
6	2507.65000000000\\
8	3343.54000000000\\
10	4179.42000000000\\
12	5015.31000000000\\
14	5851.19000000000\\
16	6687.08000000000\\
18	7522.96000000000\\
20	8358.85000000000\\
22	9194.73000000000\\
24	10030.6200000000\\
26	10866.5000000000\\
28	11702.3900000000\\
30	12538.2700000000\\
40	16717.7000000000\\
50	20897.1200000000\\
60	25076.5500000000\\
70	29255.9700000000\\
80	33435.4000000000\\
90	37614.8200000000\\
100	41794.2500000000\\
110	45973.6700000000\\
120	50153.1000000000\\
130	54332.5200000000\\
140	58511.9500000000\\
150	62691.3700000000\\
160	66870.8000000000\\
170	71050.2200000000\\
180	75229.6500000000\\
190	79409.0700000000\\
200	83588.5000000000\\
};

\end{axis}

\end{tikzpicture}%
        \end{subfigure}
        \caption{}
        \label{fig:multi_hop_eval}
    \end{subfigure}
\caption{(a) \textbf{Schematic illustration:} A repeater chain consisting of two repeaters, $R_1$ and $R_2$, connecting two end nodes, $Q_1$ and $Q_2$. Each node, i.e., repeater or end node, contains a total number of $N$ memories. Each repeater has the architecture considered in this work from~\cref{fig:Model_b} and optimally assigns memories to its sides using Lem.~\ref{lemm:optimal_alloc}.
(b) \textbf{Numerical evaluation}: Comparison of the expected fidelity and expected rate of end-to-end entanglements between (i) optimal memory allocation, (ii) fixed equal allocation and (iii) standard repeaters. The initial fidelity $F_0=1$, coherence time $t_c=0.001 \text{ sec}$ and distances $d_1=d_3=20\text{ km}$ and $d_2=30\text{ km}$.}
\label{fig:multi_node}
\vspace{-15pt}
\end{figure}

In this section, we discuss the application of the optimal memory allocation approach in repeater chains.
In repeater chains, two end nodes are connected through several repeaters as illustrated in~\cref{fig:multi_hop_model}, which depicts a two-repeater chain.
An additional aspect to be considered for multiple repeaters is their scheduling. 
Generally, scheduling protocols entail the order in which repeaters connect their adjacent nodes to eventually form an end-to-end entanglement.
Other than greedy protocols~\cite{inesta2023optimal} where all repeaters work simultaneously, other scheduling protocols may be beneficial for heterogeneous systems where repeaters and their adjacent links have different properties.
For such systems, several works propose scheduling protocols in the form of swapping trees that dictate the optimal order of repeaters given their individual achievable rates and fidelities~\cite{Ghaderibaneh,dai2020optimal,he2024parallel}.

An advantage of our optimal allocation approach is that each repeater only requires local knowledge of the entanglements on its sides.
Generally, this allows its direct application in repeater chains under any arbitrary scheduling protocol, where in each round, every repeater locally allocates the optimal number of memories to its sides.
In addition, scheduling protocols can directly leverage the theoretical bounds derived in this work, which determine the locally achievable rate and fidelity of individual repeaters, to optimize the swapping order. 

For further illustration, we apply the optimal memory allocation to the two-repeater chain in~\cref{fig:multi_hop_model} under a greedy protocol.
Specifically, in the entanglement generation phase, every two nodes simultaneously attempt to generate entanglements.
In the matching phase, the repeaters apply swapping only when an entanglement exists between every two nodes. 
For example, if the two entanglements: $R_1-Q_1$ and $R_1-R_2$ exist, they will wait until an $R_2-Q_2$ entanglement is generated in the next round, and then apply entanglement swapping.
After each round, each repeater optimally assigns memories to its left and right sides using Lem.~\ref{lemm:optimal_alloc}.
An important detail is that the two connected sides of adjacent repeaters may be assigned different numbers of memories. 
In that case, the excess memories become idle, i.e., do not attempt to generate entanglements, resulting in a loss in rate.

To demonstrate that, we numerically evaluate in~\cref{fig:multi_hop_eval} the two-repeater system from~\cref{fig:multi_hop_model} with equal total number of memories, distances $d_1=d_3=20~\text{km}$ and $d_2=30~\text{km}$, initial fidelity $F_0=1$ and coherence time $t_c=0.001~\text{sec}$.
Consistent with the single repeater evaluation in~\cref{sec: eval},~\cref{fig:multi_hop_eval} demonstrates the significant fidelity improvement of the optimal allocation and the detrimental impact of equal memory allocation.
Further, the figure shows a slightly larger loss in the rate of the repeater chain with respect to the standard repeater rate, compared to the single repeater system (c.f.~\cref{fig:rates_1}). 
Note that for the standard repeater evaluation, we let the two repeaters form an entanglement ($R_1-Q_1$) and ($R_2-Q_2$) with their adjacent quantum nodes, then form entanglements ($R_1-R_2$) between each other.
While other possible scheduling protocols for entanglement generation in the standard repeater chain may lead to relatively better fidelity and/or rate results, the overall behavior is not expected to be significantly different.

Overall, the previous example demonstrates the advantage of applying optimal allocation in repeater chains, while we leave further analysis of the impact on a large number of repeaters and under different scheduling protocols for future work.
In addition, based on the previous analysis, an important open problem is the underutilization of repeater memories.
In particular, the difference in the allocated memories between the two connected sides of adjacent repeaters results in idle memories.
A possible solution to mitigate this issue, particularly in heterogeneous systems, is to optimize the total number of memories at each repeater to minimize the number of idle memories (e.g., in terms of the mean). 
This problem becomes even more interesting when analyzed under different scheduling protocols.

\vspace{-5pt}
\section{Conclusion}
\vspace{-5pt}
We considered an asymmetric quantum repeater that simultaneously generates entanglements with its adjacent nodes located at different distances from the repeater.
We derived the optimal memory allocation that minimizes the expected number of unmatched entanglements.
Further, we derived performance bounds on the fidelity and the rate given this optimal allocation.
For specific repeater parameters, we derived the threshold on the number of unmatched entanglements beyond which the optimal allocation condition is violated.
To this end, we proposed a hard-cutoff allocation regime that drops the excess unmatched entanglements. 
Numerical evaluations showed that for arbitrary repeater parameters, the optimal allocation yields significantly better fidelity than the standard repeater model at similar rate. 
Finally, we showed that the optimal allocation dominates the fixed allocation, which may even perform worse than the standard repeater model. 

\vspace{-5pt}
\section{Appendix}
\vspace{-5pt}
\subsection{Expected number of matched entanglements}
\label{app:asymmetric_rate}
We derive the stationary expected number of matched entanglements $\Ebb[M]$ in terms of the expected unmatched entanglements.
We start from the expectation in~\cref{eq:E_matched}
\vspace{-5pt}
\begin{equation}
\label{eq:E_matched_app}
    \Ebb[M] =\frac{1}{2}\left(\Ebb[X_l]+\Ebb[X_r]\right)\,.
\vspace{-5pt}
\end{equation}
From~\cref{eq:ent_gen_binom}, the number of successful entanglements $X_l(t)$ and $X_r(t)$ are binomially distributed, where the conditional expectations given the unmatched entanglements $\alpha(t)$ are
\vspace{-5pt}
\begin{align}
    \Ebb[X_l(t)|\alpha(t)]=p_l(N_l(t)  -\alpha(t) 1_{\{\alpha(t) \geq 0\}}) \nonumber \,, \\ 
    \Ebb[X_r(t)|\alpha(t)]=p_r(N_r(t)+  \alpha(t) 1_{\{\alpha(t)<0\}}) \,,
    \label{eq: cond_exp_lr}
\vspace{-5pt}
\end{align}
where the function $1_{\{.\}}$ is the indicator function.
As the bank sizes $N_l(t)$ and $N_r(t)$ depend on $\alpha(t)$ under the optimal memory allocation, by taking the expectation over $\alpha(t)$, we obtain under stationarity
\vspace{-5pt}
\begin{align}
\label{eq:E_X}
    \Ebb[X_l]=p_l(\Ebb[N_l]  -\Ebb[\alpha \, 1_{\{\alpha \geq 0\}}]) \nonumber \,, \\ 
    \Ebb[X_r]=p_r(\Ebb[N_r]+  \Ebb[\alpha \, 1_{\{\alpha<0\}}]) \,.
\vspace{-5pt}
\end{align}
By inserting these expectations into~\cref{eq:E_matched_app}, we derive the expected number of matched entanglements as
\vspace{-5pt}
\begin{align}
     \Ebb[M] &=\frac{1}{2}\bigg(p_l\Ebb[N_l] +p_r \Ebb[N_r] \nonumber \\
     &-
     \left(p_l \Ebb[\abs{\alpha} \, 1_{\{\alpha \geq 0\}}])+ p_r\Ebb[\abs{\alpha} \, 1_{\{\alpha<0\}}]\right)\bigg)\,.
\vspace{-5pt}
 \end{align}
Here, we use the fact that $\Ebb[\alpha \, 1_{\{\alpha \geq 0\}}]=\Ebb[\abs{\alpha} \, 1_{\{\alpha \geq 0\}}]$ and {$\Ebb[\alpha \, 1_{\{\alpha<0\}}]=-\Ebb[\abs{\alpha} \, 1_{\{\alpha<0\}}]$}. 

\vspace{-5pt}
\subsection{Proof of Lem.~\ref{lemm:optimal_alloc}}
\label{app:optimal_alloc}
We derive the left and right bank sizes $N_l(t)$ and $N_r(t)$ at round $t$, which drives the conditional expectation of the unmatched entanglements $\alpha(t+1)$ to zero, i.e., $\Ebb[\alpha(t+1)|\alpha(t)]=0$.
The conditional expectation of the unmatched entanglements at round $t+1$ from~\cref{eq:mismatch_t} is
\vspace{-5pt}
\[\Ebb[\alpha(t+1)|\alpha(t)]=\alpha(t)+\Ebb[X_l|\alpha(t)]-\Ebb[X_r|\alpha(t)]\,.
\vspace{-5pt}
\] 
Using the expressions of $\Ebb[X_l|\alpha(t)]$ and $\Ebb[X_r|\alpha(t)]$ from~\cref{eq: cond_exp_lr}, we obtain
\vspace{-5pt}
\begin{align*}
    \Ebb[\alpha(t+1)|\alpha(t)]&=p_lN_l(t)-p_r N_r(t) +\alpha(t) \nonumber \\
    &-\alpha(t) (1_{\{\alpha(t) \geq 0\}}p_l+1_{\{\alpha(t) < 0\}}p_r) \nonumber \\
    &=p_lN_l(t)-p_r N_r(t) +\alpha(t)(1-g(\alpha(t))\,,
\vspace{-5pt}
    \end{align*}
where $g(\alpha(t))=1_{\{\alpha(t) \geq 0\}}p_l+1_{\{\alpha(t) < 0\}}p_r$.
Since the total number of memories $N_r(t)+N_l(t)=N$, we derive the size of the right bank resulting in $  \Ebb[\alpha(t+1)|\alpha(t)]=0$ as
\vspace{-5pt}
\begin{equation}
    p_l(N-N_r(t))-p_r N_r(t) +\alpha(t)(1-g(\alpha(t))=0 \, \nonumber \\
\vspace{-5pt}
\end{equation}
which yields the allocation in~Lem.~\ref{lemm:optimal_alloc}.

\vspace{-5pt}
\subsection{Proof of Lem.~\ref{Lemm:rate_bound}}
\label{app:rate_bound}
We derive the lower bound $E[M]\geq\mathcal{B}$ of the expected number of matched entanglements per repeater round under the optimal allocation condition $ \Ebb[\alpha(t+1)|\alpha(t)]=0$.
We start from the lower bound in terms of $\Ebb[\abs{\alpha}]$ given in~\cref{eq:n_ent_bound_general} as
\vspace{-5pt}
\[
\Ebb[M]\geq\frac{1}{2}\left( p_l\Ebb[N_l]+ p_r\Ebb[N_r]- \max{(p_l,p_r)} \Ebb[\abs{\alpha}]\right)\,,
\vspace{-5pt}
\]
Using Cauchy–Schwarz inequality, the expectation  $\Ebb[\abs{\alpha}]$ is bounded as $\Ebb[\abs{\alpha}]\leq  \sqrt{\Ebb[\alpha^2]}$.
Inserting this inequality in the matched entanglement bound, we obtain
\begin{equation}
\label{eq:M_bound}
\Ebb[M]\geq\frac{1}{2}\left(\Ebb[N_l] p_l+\Ebb[N_r] p_r- \max{(p_l,p_r)} \sqrt{\Ebb[\alpha^2]}\right)\,.
\end{equation}
Now we derive the second moment ${\Ebb[\alpha^2]}$. 
Under stationarity, the memory allocation condition $\Ebb[\alpha(t+1)|\alpha(t)]=0$ implies that the marginal expectation $\Ebb[\alpha]=0$. 
Hence, the second moment ${\Ebb[\alpha^2]}=\text{Var}[\alpha]$.
From~\cref{eq:mismatch_t}, the non-stationary conditional variance
\begin{equation}
\label{eq:var_1}
    \text{Var}[\alpha(t+1)|\alpha(t)]=\text{Var}[X_l(t)|\alpha(t)]+\text{Var}[X_r(t)|\alpha(t)]\,.
\end{equation}
Recall that $X_l(t)$ and $X_r(t)$ are the number of successful entanglements, which are binomially distributed as defined in~\cref{eq:ent_gen_binom}.
Accordingly,
\begin{align*}
    \text{Var}[X_l(t)|\alpha(t)]=p_l(1-p_l)( N_l(t) -\alpha(t) 1_{\{\alpha(t) \geq 0\}}) \nonumber \,, \\ 
    \text{Var}[X_r(t)|\alpha(t)]=p_r(1-p_r)(N_r(t)+  \alpha(t) 1_{\{\alpha(t)<0\}}) \,.
\end{align*}
Using these expressions in~\cref{eq:var_1}, we obtain
\begin{align}
      \text{Var}[\alpha(t+1)|\alpha(t)]&=p_l(1-p_l)N_l(t)+p_r(1-p_r)N_r(t) \nonumber \\
      &-\abs{\alpha(t)}h(\alpha(t))\,,
\end{align}
where $h(\alpha(t))= 1_{\{\alpha(t) \geq 0\}} p_l(1-p_l)+1_{\{\alpha(t) < 0\}}p_r(1-p_r)$.
As the value of the term $\abs{\alpha(t)}h(\alpha(t))\geq 0$, the conditional variance is bounded as
\vspace{-5pt}
\[
 \text{Var}[\alpha(t+1)|\alpha(t)]\leq p_l(1-p_l)N_l(t)+p_r(1-p_r)N_r(t) \,.
\]
Taking the expectation over $\alpha(t)$, we obtain under stationarity
\vspace{-5pt}
\begin{equation}
\label{eq:var_alph}
\text{Var}[\alpha]\leq p_l(1-p_l)\Ebb[N_l]+p_r(1-p_r) \Ebb[N_r]\,.
\vspace{-5pt}
\end{equation}

Next, we derive the stationary expectations of the memory bank sizes, i.e, $\Ebb[N_l]$ and $\Ebb[N_r]$, using the memory allocation from Lem.~\ref{lemm:optimal_alloc} as 
\vspace{-5pt}
\begin{align}
\label{eq:E_N}
    \Ebb[N_r]&=({p_l}/({p_l+p_r}))\left(N+{\Ebb[\alpha (1-g(\alpha))]}
    /{p_l}\right) \nonumber \,, \nonumber \\
    \Ebb[N_r]&\approx{p_l N}/({p_l+p_r})\,, \;  \Ebb[N_l]\approx{p_r N}/({p_l+p_r}) \,,
\vspace{-5pt}
\end{align}
where $g(\alpha(t))=1_{\{\alpha(t) \geq 0\}}p_l+1_{\{\alpha(t) < 0\}}p_r$.
Here in the first line, we use the original expression on $N_r(t)$ before rounding.
Observe in the second line of the equation that we approximate the expressions since the value of $\Ebb[\alpha (1-g(\alpha))]/p_l$ is  small under mismatch minimization.
From Lem.~\ref{lemm:optimal_alloc} the mean $\Ebb[\alpha]=0$  and consequently, $\Ebb[\alpha \,  1_{\{\alpha \geq 0\}}]=-\Ebb[\alpha \, 1_{\{\alpha<0\}}]= \frac{1}{2} \Ebb[\abs{\alpha}]$.
Hence, we obtain $\Ebb[\alpha (1-g(\alpha))]=(p_r-p_l) \Ebb[\abs{\alpha}]/2$. 
As a result, the value of $\Ebb[\alpha (1-g(\alpha))]/p_l$ is small, particularly compared to the total number of memories $N$, since the expected mismatch, i.e., $\Ebb[\abs{\alpha}]$, is minimized.
Recall that the memory allocation under Lem.~\ref{lemm:optimal_alloc} minimizes the expected mismatch under the assumption that the mean and the median of $\alpha$ coincide.
Note that an implicit assumption underlying the above approximation is that there is a sufficient total number of memories that satisfies the allocation in Lem.~\ref{lemm:optimal_alloc}, avoiding its violation as discussed in~\cref{sec:Alloc_viol}.
This assumption holds asymptotically.
Finally, using~\cref{eq:E_N} together with~\cref{eq:var_alph} in~\cref{eq:M_bound}, we obtain the bound in Lem.~\ref{Lemm:rate_bound}.

\balance
\bibliographystyle{IEEEtran.bst}
\bibliography{IEEEabrv,bibliocache.bib}

\end{document}